\documentclass[epj]{svjour}
\usepackage{times,mathptmx}
\usepackage{graphicx}
\usepackage[numbers,sort&compress]{natbib}

\begin{document}

\title{Finite temperature properties and frustrated ferromagnetism
       in a square lattice Heisenberg model}

\author{Nic Shannon\inst{1} \and Burkhard Schmidt\inst{2} 
\and Karlo Penc\inst{3} \and Peter Thalmeier\inst{2}}

\institute{SPEC, CEA Saclay, Orme des Merisiers
F-91191 Gif sur Yvette CEDEX, France. 
\and 
Max-Planck-Institut f{\"u}r
Chemische Physik fester Stoffe,
N{\"o}thnitzer Str. 40, 01187 Dresden, Germany.
\and 
Research Institute of Solid State Physics and Optics,
H-1525 Budapest, P.O.B. 49, Hungary}

\abstract{The spin 1/2 Heisenberg model on a square lattice with
antiferromagnetic nearest- and next-nearest neighbour interactions (the
$J_1$--$J_2$ model) has long been studied as a paradigm of a
two-dimensional frustrated quantum magnet.  Only very recently,
however, have the first experimental realisations of such systems been
synthesized.  The newest material, Pb$_2$VO(PO$_4$)$_2$ seems to have
mixed ferro-- and antiferromagnetic exchange couplings.  In the light of
this, we extend the semiclassical treatment of the $J_1$--$J_2$
model to include ferromagnetic interactions, and present an analysis of
the finite temperature properties of the model based on the exact
diagonalization of 8, 16 and 20 site clusters.  We propose that diffuse
neutron scattering can be used to resolve the ambiguity inherent in
determining the ratio and sign of $J_1$ and $J_2$ from thermodynamic
properties alone, and use a finite temperature Lanczos algorithm to
make predictions for the relevant high temperature spin-spin
correlation functions.  The possibility of a spin-liquid phase
occurring for ferromagnetic $J_1$ is also briefly discussed.
\PACS{ {71.27.+a}{Strongly correlated electron systems; heavy fermions}
\and {71.10.-w}{Theory and models of many-electron systems} \and
{75.40.Cx}{Static properties (order parameter, static susceptibility,
heat capacities, critical exponents, etc.)}}}

\date{Received: date / Revised version: date}

\maketitle

\section{Introduction}

The antiferromagnetic (AF) Heisenberg model on a square lattice with
added nearest neighbour bonds, commonly referred to as the $J_1$-$J_2$
model, has long served as a paradigm for a two dimensional frustrated
AF. For a model with only one adjustable parameter -- the ratio of
next-nearest to nearest neighbour exchange $J_2/J_1$ -- it has an
extremely rich phase diagram, offering the chance to study N\'eel order
(NAF) with a reduced sublattice magnetization for $J_2/J_1 \ll 1$, a
collinear AF (CAF) phase selected by an order from disorder effect for
$J_2/J_1 \gg 1$, and a spin-gapped phase (or family of phases) for
intermediate coupling.  A correspondingly rich and occasionally
controversial literature, has accompanied the development of these
ideas~\cite{misguich:03-1}.

More recently, the discovery of high T$_c$ superconductivity in doped
layered cuprates whose undoped parent compounds are spin-half
antiferromagnets, has lead to a renaissance of interest in 2D
frustrated magnets.  The $J_1$-$J_2$ model in particular has attracted
renewed attention, both because of its simplicity and the possibility
that its spin-gapped phase might provide a realization of Anderson's
resonating valence bond (RVB) concept~(see e.g.~\cite{sorella:98}).

Given all of this theoretical activity, it is surprising that the behaviour of the 
$J_1$-$J_2$ model in the presence of ferromagnetic (FM) exchange remains 
largely unexplored.

Frustrated FM's have an interesting history in their own right.  The
solid phases of He~III have very complex magnetic behaviour determined
by competing FM and AF multiple spin exchange processes.  Under
appropriate conditions He~III may exhibit FM, AF or spin liquid ground
states.  Another, solid state, example is provided by the doped
colossal magnetoresistance (CMR) manganites, where the competition
between kinetic energy driven FM and superexchange driven AF is widely
believed to result in a phase separation into regions with different
magnetic order.  It is therefore worth asking whether the $J_1$-$J_2$
model with FM $J_1$ shows similarly exotic behaviour.

It is also perhaps a little surprising, considering the wide variety of
magnetic materials now under study, that the first ``$J_1$-$J_2$
compound'', Li$_2$VOSiO$_4$, was discovered only very recently
\cite{millet:98,melzi:00,melzi:01}.  Li$_2$VOSiO$_4$ is an insulating
Vanadium oxide, with spin $S=1/2$ V$^{4+}$ ions arranged in square
lattice planes, at the centres of VO$_4$ pyramids.  These are linked by
SiO$_4$ tetrahedra, with Li ions occupying the space between the V--O
planes.  Because of this relatively complex structure, the magnetic
ions are well separated, with weak superexchange between nearest and
next-nearest neighbour V ions mediated by more than one intermediate O
ion.  A small number of related materials have now been synthesized,
and the preliminary analysis of one of these, Pb$_2$VO(PO$_4$)$_2$
seems to offer evidence of mixed FM and AF exchange coupling
\cite{kaul:03-1}.  However, a precise and unambiguous measurement of the
exchange couplings $J_1$ and $J_2$ by, e.\,g., inelastic neutron
scattering, has yet to be accomplished for {\it any} of these new
compounds.

Motivated by Pb$_2$VO(PO$_4$)$_2$, in this paper we provide an
overview of the ground state  and finite temperature properties of the
$J_1$-$J_2$ with mixed FM and AF couplings.  We present a comprehensive
semiclassical analysis of the three dominant ordered phases of the
model -- a uniform FM phase, and $\vec{q} = (\pi,\pi)$ N\'eel (NAF) and
$\vec{q}^* = (\pi,0), (0,\pi)$ collinear (CAF) antiferromagnetic
phases -- together with an exact analytic diagonalization of an eight
site cluster, and finite temperature Lanczos results for the heat
capacity and magnetic susceptibility for 16 and 20 site clusters.

We argue that, in addition to the known spin liquid region for $J_1 >
0$, $J_2 \sim J_1/2$, where the NAF and CAF phases compete, a new spin
liquid region may exist for $J_1 < 0$, $J_2 \sim -J_1/2$, where the FM
and CAF phases compete.  We also propose that, because of their low
magnetic energy scales, diffuse neutron scattering at finite
temperatures can provide a very useful source of information about the
nature of the competing magnetic interactions in these materials.  With
this in mind, we present the first quantitative numerical estimates of
the magnetic structure factor S($\vec{q}$, T) for the $J_1$-$J_2$
model.

\section{Zero temperature properties}

\subsection{Classical phase diagram and general arguments}
\label{general}

We consider the spin 1/2 Heisenberg model on a square lattice
\begin{equation}
{\cal H} = J_1 \sum_{\langle ij \rangle_1} \vec{S}_i\cdot\vec{S}_j
	  + J_2 \sum_{\langle ik \rangle_2} \vec{S}_i\cdot\vec{S}_k
\label{eqn:H}
\end{equation}
where the sum on $\langle ij \rangle_1$ runs over nearest neighbour
and the sum $\langle ik \rangle_2$ over diagonal next-nearest neighbour 
bonds.  We allow the exchange constants $J_1$ and $J_2$ to be negative
(FM) as well as positive (AF) 
%(for an illustration see Figure~\ref{fig:fig0}).

Since the properties of the $J_1$-$J_2$ model depend on the 
relative and not the absolute size of the exchange couplings
$J_1$ and $J_2$, it is convenient to
characterize it by an overall energy scale 
\begin{equation} 
J_{\rm c} = \sqrt{J_1^2 + J_2^2}
\end{equation}
and a frustration angle $\phi$ such that 
\begin{eqnarray} 
J_1 &=& J_{\rm c} \cos(\phi)\qquad J_2 = J_{\rm c} \sin(\phi)\nonumber\\
\phi&=&\tan^{-1}(J_2/J_1)
\end{eqnarray}
As discussed in Section~\ref{finiteT}, $J_{\rm c}$ can in principle be determined 
directly from the asymptotic behaviour of heat capacity
and susceptibility at high temperatures.  However the different
physical properties of the model depend on the angle $\phi$,
and this is much harder to determine from experiment.

Let us first consider the simplest possible classical analysis of the
model.  We assume that the system orders at zero temperature in such a
way that all the spins are oriented in a common
plane~\cite{villain:59}.  In this case the ground state energy of the
$J_1$-$J_2$ model is minimised by an order parameter with wave vector
$\vec{q}=\vec{q}^*$ such that the energy per spin 
\begin{equation}
    \label{eqn:classicalE}
    E(\vec{q}^*) =  \frac{1}{2} z S^2
    \left[J_1 \gamma_1(\vec{q}^*) + J_2 \gamma_2(\vec{q}^*)\right]
\end{equation}
takes on its minimal value.  Here
\begin{eqnarray}
\gamma_1(\vec{q}) &=& (\cos(q_x) + \cos(q_y))/2 \nonumber\\
\gamma_2(\vec{q}) &=& \cos(q_x) \cos(q_y)
\end{eqnarray}
$z=4$ is the lattice coordination number for each type of bond 
and $S=1/2$ is the size of the spin.

\begin{figure}
    \begin{center}
	\includegraphics[width=.8\columnwidth]{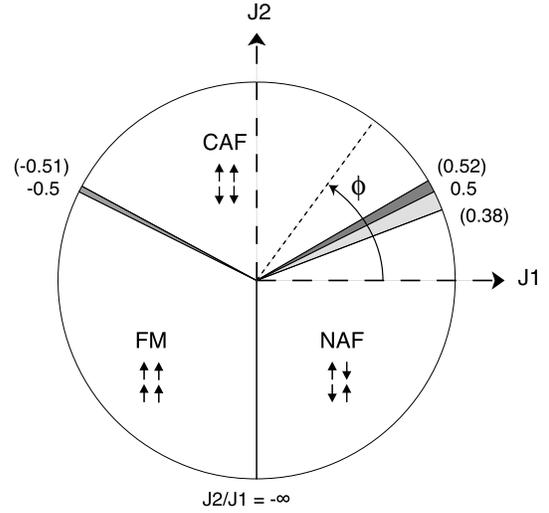}	
    \end{center}
    \caption{Classical phase diagram.  Numbers are ratios of exchange
    couplings $J_2/J_1$ for phase boundaries as determined from the
    classical ground state energy.  The boundary between FM and NAF
    phase is the line $J_{1}=0$, $J_{2}<0$ ($J_{2}/J_{1}=-\infty$ in
    the figure).  Values of $J_{2}/J_{1}$ in parentheses show where
    zero point fluctuations destroy the relevant order parameter at a
    semiclassical level, as discussed in Section~\ref{spinwave}.  The
    shaded areas for $J_{1}>0$ correspond to the known spin-liquid
    regime, and for $J_{1}<0$ to another spin liquid region.  The
    frustration angle is given by $\phi=\tan^{-1}(J_{2}/J_{1})$.}
    \label{fig:phasediag}
\end{figure}
\begin{table*}[tb]
    \begin{center}
	\begin{tabular}{lllll} 
	    \hline\noalign{\smallskip}
	    &
	    $\vec{q}^*$ 
	    & Energy
	    & Range ($J_1$,$J_2$) 
	    & Range $\phi$ \\  
	    \noalign{\smallskip}\hline\noalign{\smallskip}
	    NAF &
	    $(\pi,\pi)$ &
	    $-J_1/2 + J_2/2$& 
	    $J_1>0, \quad J_2 < J_1/2 $ & 
	    $ -\pi/2 < \phi < \tan^{-1}(\frac{1}{2})$ \\
	    \noalign{\smallskip}
	    \noalign{\smallskip}
	    CAF &
	    $(0,\pi)$ or $(\pi,0)$ &
	    $-J_2/2$& 
	    $\mid J_2 \mid > \mid J_1 \mid /2$ & 
	    $\tan^{-1}(\frac{1}{2}) < \phi < \pi - \tan^{-1}(\frac{1}{2}) $ \\   
	    \noalign{\smallskip}
	    \noalign{\smallskip}
	    FM &
	    $(0,0)$ &
	    $J_1/2 + J_2/2$ &
	    $J_1 < 0, \quad J_2 < -J_1/2$ & 
	    $ \pi -\tan^{-1}(\frac{1}{2}) < \phi < -\pi/2 $  \\   
	    \noalign{\smallskip}\hline
	\end{tabular}
	\caption{ Parameters for classical ground states diagram.  }
	\label{table1}
    \end{center}
\end{table*}
\begin{figure}
    \begin{center}
	\includegraphics[width=.8\columnwidth]{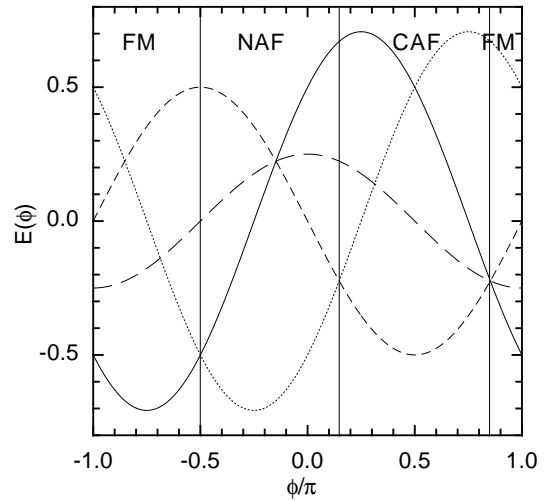}

	\caption{Classical energies $E(\vec{q}^*)$ as a function of the
	frustration angle $\phi$ in units of $J_{\rm c}$.  Solid line:
	FM, $\vec{q}^*=0$; dotted line: NAF, $\vec{q}^*=(\pi,\pi)$;
	dashed line: CAF, $\vec{q}^*=(\pi,0)$; long-dashed line: four
	sublattice state for $\vec{q}^* = (\pi/2, 0)$.}
	\label{fig:Eclassical}
    \end{center}
\end{figure}
This analysis selects three phases, a N\'eel AF (NAF), a collinear AF
(CAF) and a uniform FM to give the phase diagram shown in
Figure~\ref{fig:phasediag} with the parameters given in
Table~\ref{table1}.  Note that the coplanar spiral states with 
$\vec{q}^* = (2\pi n/m,0)$, where $\{n,m\}$ are integers, have energies 
which interpolate between the CAF and FM, and {\it all} of these 
states become degenerate exactly
at the transition from FM to CAF. The classical energies of FM, NAF and
CAF order parameters, together with a four sublattice state with
$\vec{q}^* = (\pi/2, 0)$ are shown in Figure~\ref{fig:Eclassical}.

Each of the three lines selected as classical phase boundaries have
interesting properties which are related to symmetries of the
Hamiltonian.  The mirror symmetry of the classical phase diagram about
the line $J_1 = 0$ is particularly easy to understand.  Since the
square lattice is bipartite, the spins can be divided into A and B
sublattices and, the Hamiltonian remains invariant under the
transformation
\begin{equation} 
\vec{S}^B \to - \vec{S}^B \qquad J_1 \to -J_1 
\end{equation}
This converts the classical NAF into a FM, 
and the classical $\vec{q}^* = (0,\pi)$ CAF into the 
classical $\vec{q}^* = (\pi,0)$ CAF.  
Exactly on the line $J_1 = 0$ the A and B sublattices 
are entirely disconnected, so it is possible to rotate the 
classical NAF into the FM, and 
the classical $\vec{q}^* = (\pi,0)$ CAF into the 
classical $\vec{q}^* = (0,\pi)$ CAF, without any cost of energy.  

Of course it is reasonable to ask whether such a naive classical
picture has any relevance at all for the physics of a two dimensional
frustrated spin 1/2 magnet.  By way of an answer, let us consider in
turn the simplest possible quantum analysis of the model.

Using the geometric trick of double counting
all $J_1$ bonds and then setting $J_1 \to J_1/2$, the  
cross-linked square lattice can be expressed
as four interpenetrating sublattices of edge sharing
tetrahedra (cross-linked squares).  And since each 
of these cross-linked squares is a complete graph, the 
Hamiltonian~(\ref{eqn:H}) can be rewritten
in terms of the total spin on a square 
\begin{equation}
\label{eqn:special}
{\cal H} = - 2J_2 S(S+1) N + \sum_{\fbox{}} {\cal H}_{\fbox{}}
\end{equation}
where the sum runs over
all $N$ squares $\fbox{}$ of the lattice, and 
\begin{equation}
\label{eqn:H4site}
{\cal H}_{\fbox{}}  = \frac{J_1}{4}\Omega_+^2
      + \frac{1}{2}\left(J_2 - \frac{J_1}{2} \right)
	\left[\Omega_A^2 +  \Omega_B^2\right]
\end{equation}
The spins within each square are counted clockwise (or
counterclockwise) such that
\begin{eqnarray}
\Omega_+ &=& \Omega_A  
   +\Omega_B \\
\Omega_A &=&  \vec{S}^A_1 + \vec{S}^A_3\\ 
\Omega_B &=& \vec{S}^B_2 + \vec{S}^B_4
\end{eqnarray}
With the Hamiltonian written in this way, the special role of the
classical transition lines $J_2 \pm J_1/2 = 0$ becomes self-evident.
Moreover, on these lines, the lowest energy is achieved by {\it any}
state for which the sum (difference) of the A and B sublattice
magnetisations vanishes {\it in each square}.

\begin{figure}
    \begin{center}
	\includegraphics[width=.8\columnwidth]{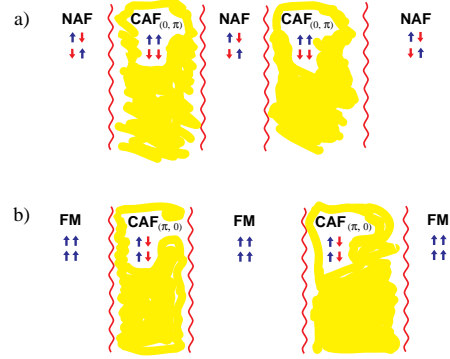}
    \end{center}
    \caption{On the transition lines a) $J_2/J_1 = 0.5$ and b) $J_2/J_1
    = -0.5$ the Ising domain wall energy between CAF and NAF or FM
    states vanishes and the system can break up into stripe like
    domains with all possible values of total magnetization.}
    \label{fig:stripes}
\end{figure}
Classically, this condition can be satisfied by an exponentially large
number of states.  As a result, at these transitions, it becomes
possible to break the system up into stripes.  In the Ising limit these
stripes are the alternating domains illustrated in
Figure~\ref{fig:stripes}.  The special (local) degeneracy of the
Hamiltonian~(\ref{eqn:special}) also reveals itself in lines of zeros
in the spinwave spectrum, discussed below.

We now use equation~(\ref{eqn:special}) to construct a minimal quantum
theory for the model.  The state of the whole system is determined if
we specify all of the spins on any of the four sublattices of
tetrahedra.  So a lower quantum bound on the ground state energy can be
obtained by considering a single isolated tetrahedron.

\begin{table*}[tb]
    \begin{center}
	\begin{tabular}{lllll} 
	    \hline\noalign{\smallskip}
	    &
	    $(\Omega,\omega_A,\omega_B)$ &
	    Energy &
	    Range ($J_1$,$J_2$) &
	    Range $\phi$ \\  
	    \noalign{\smallskip}\hline\noalign{\smallskip}
	    ``NAF'' &
	    (0,1,1) &
	    $-J_1 + J_2/2$& 
	    $J_1>0, \quad J_2 < J_1/2 $ & 
	    $ -\pi/2 < \phi < \tan^{-1}(\frac{1}{2})$ \\
	    \noalign{\smallskip}
	    \noalign{\smallskip}
	    ``CAF'' &
	    (0,0,0) &
	    $-3J_2/2$& 
	    $J_{2}>J_{1}/2$& 
	    $\tan^{-1}(\frac{1}{2}) < \phi < \pi - \tan^{-1}(\frac{1}{4}) $ \\   
	    \noalign{\smallskip}
	    \noalign{\smallskip}
	    FM &
	    (2,1,1) &
	    $J_1/2 + J_2/2$&
	    $J_1 < 0, \quad J_2 < -J_1/4$ & 
	    $ \pi -\tan^{-1}(\frac{1}{4}) < \phi < -\pi/2 $  \\   
	    \noalign{\smallskip}\hline
	\end{tabular}
    \end{center}
    \caption{Parameters for the ground states of a single cross linked
    square.}
    \label{table2}
\end{table*}
\begin{figure}
    \begin{center}
	\includegraphics[width=.8\columnwidth]{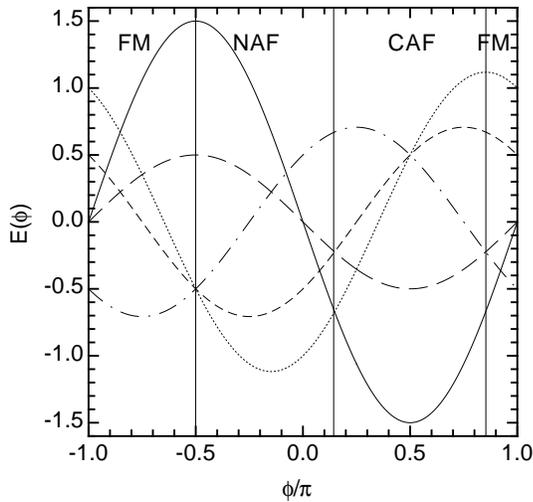}
    \end{center}
    \caption{Energies of the six eigenstates of a single tetrahedron,
    as a function of the frustration angle $\phi$, in units of $J_{\rm
    c}$.}
    \label{fig:Etetrahedron}
\end{figure}
We find what are essentially the same three ground states as a function
of $\phi$, with energies and quantum numbers given in
Table~\ref{table2}.  The full set of energy eigenvalues is shown in
Figure~\ref{fig:Etetrahedron}.  Apart from in the FM phase, where they
must agree, the upper bound on the ground state energy obtained from the
simple classical analysis, and the lower bound obtained from this
minimal quantum estimate are generally quite different.  However, it is
far more significant that the same three phases are found\footnote{Here
``phase'' should be understood to mean a ground state wave function
with the same total spin and spatial symmetries as the corresponding
classical order parameter --- SU(2) invariance clearly cannot be
spontaneously broken in a small cluster.}, with the same phase
boundaries -- excepting that between the CAF and FM, where quantum
fluctuations extend the CAF regime at the expense of the FM regime.

This simple correspondence between the most naive classical and quantum
theories lends us some confidence in both.  And although neither can
give a complete description of the model, both can be improved
systematically, as discussed in later sections of the paper.  The key
question which remains is how the phase transitions between the three
dominant phases take place.  The existence of an extensive set
of spiral states degenerate with the FM and CAF order parameters 
at $\phi = \pi - \tan^{-1}(1/2)$, and the complicated level crossings 
for the tetrahedron near $\phi = \tan^{-1}(1/2)$ and $\phi = \pi -
\tan^{-1}(1/2)$ already hint that these can be non-trivial.

\subsection{Semiclassical spin wave analysis}
\label{spinwave}

The first step to improving on the naive classical phase diagram
is to consider the influence of semiclassical spin wave excitations.
While for the FM, spin waves are eigenstates, the frustrated 
NAF and CAF phases both show marked fluctuations at zero temperature.
All three phases must be unstable at finite temperatures,
as long range order would violate the Mermin-Wagner theorem.
None the less, linear spin wave theory captures the essential
physics of the ordered phases, and provides some interesting
hints about how the classical phase diagram must be modified
in the quantum case.

\subsubsection{FM phase}

\begin{figure}
    \begin{center}
	\includegraphics[width=.8\columnwidth]{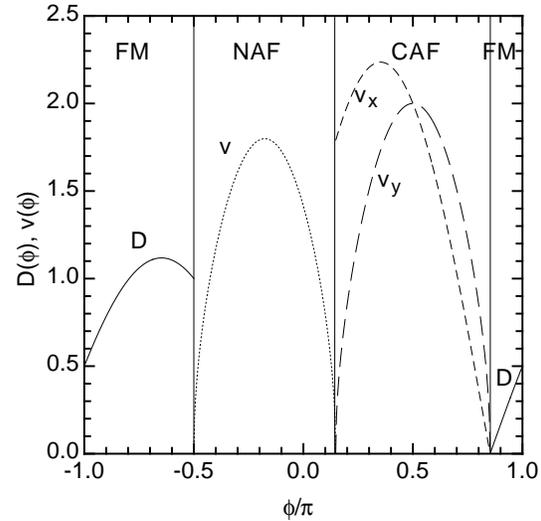}
    \end{center}
    \caption{Evolution of the spin stiffness $D$ in the FM phase (solid
    line), and of the spinwave velocities $v$ in the NAF phase (dotted
    line) and $v_x$, $v_y$ in the CAF phases (dashed lines), as a
    function of the frustration angle $\phi$ in units such that the
    lattice constant $a=1$ and $J_{\rm c}=1$.}
    \label{fig:D}
\end{figure}
\begin{figure*}
    \begin{center}
	\includegraphics[width=.8\textwidth]{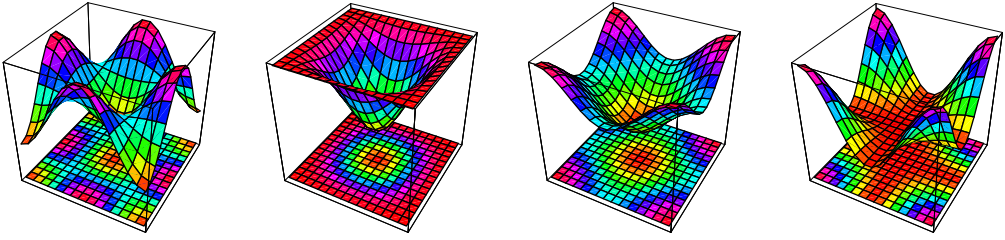}
    \end{center}
    \caption{Evolution of spinwave dispersion in FM phase.  From left
    to right -- border with NAF, deep within FM phase, pure nearest
    neighbour exchange, border with CAF.}
    \label{fig:FM}
\end{figure*}

Expanding about the FM phase we find a spin wave dispersion
\begin{equation}
\omega(\vec{q}) = -4S[J_1 + J_2] + 4S[J_1(c_x + c_y)/2 + J_2c_xc_y]
\end{equation}
where
\begin{equation} 
    c_x = \cos(q_x) \quad c_y = \cos(q_y)
\end{equation}
in units such that the lattice constant $a=1$.  The spin wave
dispersion for a range of values of $\phi$ throughout the FM phase is
shown in Figure~\ref{fig:FM}.  Note that in this and all subsequent
plots of spin wave dispersion, the $q_x$ and $q_y$ values run from
$-\pi$ to $\pi$, i.\,e.\ over the full Brillouin zone (BZ) for the
square lattice, and {\it not} the reduced magnetic BZ's appropriate to
the NAF or CAF phases.

While the fully polarized FM ground state remains an exact eigenstate of
the frustrated model, its dispersion is profoundly modified by
competing interactions.  At the boundary with the NAF phase for $\phi_c
= -\pi/2$, the dispersion is that of a pure $J_2$ FM, which has the
same magnetic BZ as the NAF phase, and therefore zeroes at $\vec{q} =
(\pi,\pi)$ and symmetry points, in addition the usual $\vec{q} = (0,0)$
Goldstone mode.  Deep within the FM phase for \mbox{$\phi = -\pi +
\tan^{-1}(1/2)$}, the dispersion behaves as
\begin{equation} 
    \omega(\vec{q}) \sim Dq^2
\end{equation}
where the stiffness constant D is given by
\begin{equation} 
    D = -(J_1 + 2 J_2)S
\end{equation}
in the zone centre, but vanishes on the zone boundary.  The variation
of $D$ as a function of $\phi$ is plotted in Figure~\ref{fig:D} For
$\phi = -\pi$ the dispersion is that of the familiar pure $J_1$ FM.
And, finally, on the boundary between FM and CAF for $\phi_c = \pi
-\tan^{-1}(1/2)$, the dispersion vanishes on the lines $q_x=0$ and
$q_y=0$.  These lines of zeros are a direct manifestation of the
special local symmetry discussed in Section~\ref{general}.

\begin{figure}
    \begin{center}
	\includegraphics[width=.8\columnwidth]{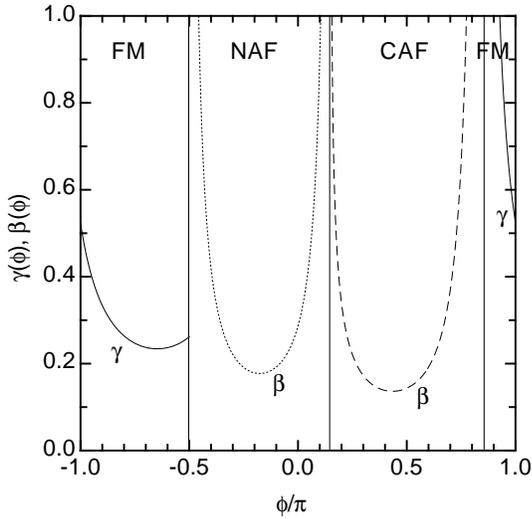}
    \end{center}
    \caption{Evolution of the heat capacity $C_V$ as a function of the
    frustration angle $\phi$.  In FM regions the quantity plotted is
    the prefactor $\gamma$ of $C_{V}=\gamma T$, and in AF regions the
    prefactor $\beta$ of $C_{V}=\beta T^2$, where temperature is
    measured in units of $J_{\rm c}$.}
    \label{fig:cV}
\end{figure}
The heat capacity of a FM in 2D is linear at low temperatures,
reflecting a constant density of states at zero energy,
and scales as 
\begin{equation}
    \label{specNAF}
    C_V = \frac{\zeta(2)}{2 \pi}\left(\frac{T}{D}\right)
\end{equation}
where $\zeta(2) = \pi^2/6$.
The coefficient of $T$ as function of $\phi$ is plotted in
Figure~\ref{fig:cV}.  It diverges at the transition between
the FM and the CAF, but approaches a
constant at the transition between FM and NAF.

\subsubsection{NAF phase}

\begin{figure*}
    \begin{center}
	\includegraphics[width=.8\textwidth]{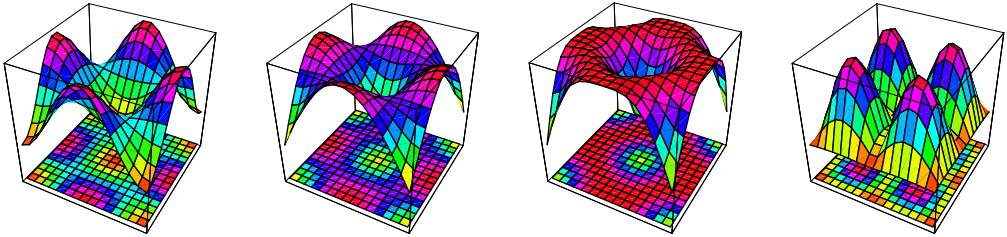}
    \end{center}
    \caption{Evolution of spinwave dispersion in NAF phase.  From left
    to right -- border with FM, deep within NAF phase, pure nearest
    neighbour exchange, border with CAF.}
    \label{fig:NAF}
\end{figure*}
The spinwave spectrum for the NAF is given by 
\begin{equation} 
    \label{eqn:AB}
    \omega(\vec{q}) = \sqrt{A_{\vec{q}}^2 - B_{\vec{q}}^2}
\end{equation}
where the coupling between spins on a given sublattice is 
\begin{equation} 
    A_{\vec{q}}  = 4S[J_1 - J_2 (1- c_x c_y)]
\end{equation}
and the coupling between the two sublattices is
\begin{equation} 
    B_{\vec{q}}  = 2 J_1 S (c_x + c_y)
\end{equation}
Where $J_2$ is FM, it acts to stabilize the NAF order, where $J_2$ is
AF, it acts to destroy it.  Once again this competition is visible in
the spin wave dispersion, as show in Figure~\ref{fig:NAF}.

At the boundary with the FM phase for $\phi_c = -\pi/2$, the dispersion
is that of a pure $J_2$ NAF, and exactly matches that of the FM on this
phase boundary.  Deep within the NAF phase for $\phi_c =
-\tan^{-1}(1/2)$, the low energy spin wave dispersion behaves as
\begin{equation} 
    \omega(\vec{q}) \sim v_s \mid \vec{q} - \vec{q}^*\mid
\end{equation}
where the isotropic spin wave velocity is given by
\begin{equation}
    v_s = 2S \sqrt{2J_1 (J_1 - 2 J_2)}
\end{equation}
and $q^* = (\pi,\pi)$, as expected.  However it exhibits a marked
dispersion about the magnetic zone boundary, as compared to the pure
$J_1$ NAF for $\phi = 0$.  Finally, on the boundary with the CAF phase
for $\phi_c = \tan^{-1}(1/2)$, the dispersion vanishes on the lines
$q_x=0$, $q_x = \pm \pi$ and $q_y=0$, $q_y = \pm \pi$.  Values of the
spinwave velocity $v_S$ are shown in Figure~\ref{fig:D}, in units in
such that the lattice spacing $a=1$ and the overall energy scale $J_{\rm c} =
1$.

The low temperature heat capacity of the NAF is controlled by the spin
wave velocity and is given by
\begin{equation}
    C_V = \frac{3\zeta(3)}{2\pi} \left(\frac{T}{v_S}\right)^2
\end{equation}
where $\zeta(3) = 1.202\ldots$.  The relevant coefficient of $T^2$, as
a function of $\phi$ is plotted in Figure~\ref{fig:cV}.

\begin{figure}
    \begin{center}
	\includegraphics[width=.8\columnwidth]{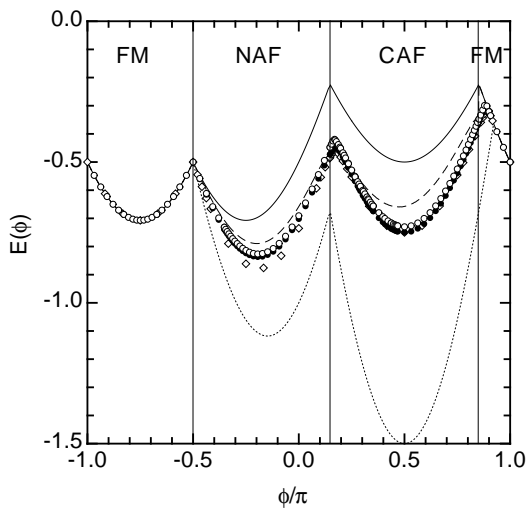}
    \end{center}
    \caption{Different estimates of the ground state energy per spin as
    a function of the frustration angle $\phi$ in units of $J_{\rm c}$.
    Uppermost (solid) line -- classical energy of FM, NAF and CAF order
    parameters given by Equation~\protect\ref{eqn:classicalE}.
    Lowermost (dotted) line -- lower quantum bound given by the ground
    state of the Hamiltonian~(\ref{eqn:special}).  Inner (dashed) line
    -- semiclassical spinwave theory.  Diamonds -- variational lower
    bound based on diagonalization of a nine-site cluster.  Solid and
    open circles -- ground state energy obtained from exact
    diagonalization of a 16-- and 20--site cluster, respectively.  Note
    that in the collinear phase the variational bound from the
    nine-site cluster and the exact-diagonalization result for the
    16-site cluster nearly coincide.}
    \label{fig:bounds}
\end{figure}
The quantum zero point corrections to the ground state energy also vary
strongly as a function of $\phi$ within the NAF state, vanishing
altogether at the boundary with the FM. These are shown in
Figure~\ref{fig:bounds}, together with the upper classical and lower
quantum bounds discussed in Section~\ref{general}.  In fact, near the
transition from NAF to CAF, the semiclassical estimate of the ground 
state energy surprisingly well with series expansion estimates \cite{singh:99}.
It also lies very close to our numerical estimates and an improved 
variational quantum bound, discussed below.  
These are also plotted in Figure~\ref{fig:bounds}.

\begin{figure}
    \begin{center}
	\includegraphics[width=.8\columnwidth]{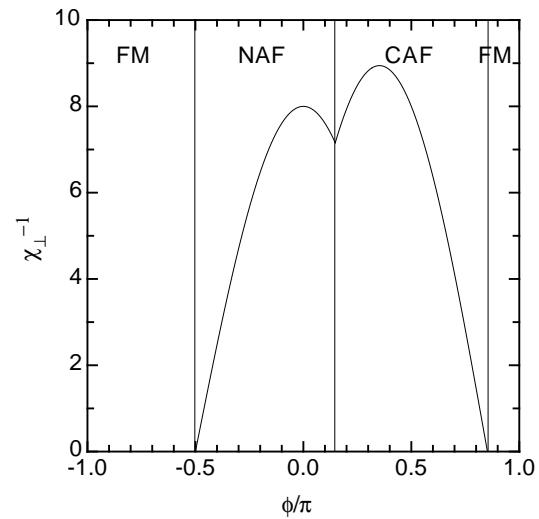}
    \end{center}
    \caption{Evolution of the inverse transverse susceptibility
    $\chi_\perp^{-1}$ as a function of the frustration angle $\phi$, in
    units such that $J_{\rm c}=1$ and $(g\mu_{\rm B})^{2}=1$.}
    \label{fig:chi}
\end{figure}
It is also interesting to consider the $\phi$ dependence of the
experimentally accessible transverse susceptibility $\chi_\perp$ which,
within spin wave theory for a two sublattice AF is given by
\begin{equation}
    \chi_\perp^{-1} = \frac{2}{S} A_{\vec{q} = \vec{0}}
\end{equation}
in units such that the overall prefactor $(g \mu_B)^2 = 1$.  The
variation of the inverse susceptibility as a function of $\phi$,
normalized to the value for $\phi = 0$, is shown in
Figure~\ref{fig:chi}.  The susceptibility is a continuous function of
$\phi$, diverging (as $\chi_\perp \sim (\phi - \phi^*)^{-1}$) at the FM
phase boundary, and matching that of the CAF at the other end of the
NAF phase.  However its derivative is discontinuous across each
transition, reflecting the changing symmetry of the order parameter.

\begin{figure}
    \begin{center}
	\includegraphics[width=.8\columnwidth]{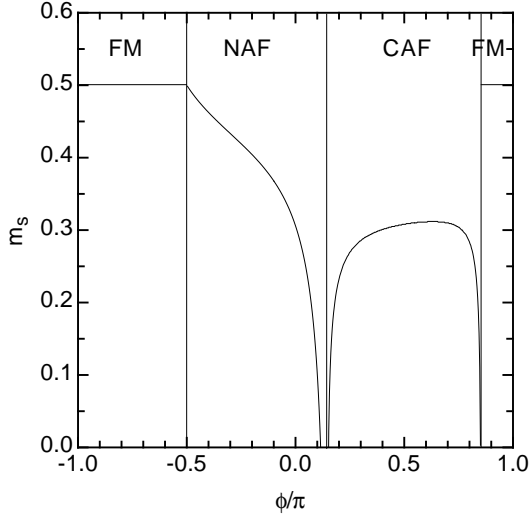}
    \end{center}
    \caption{Evolution of sublattice magnetization $m_s$ as a function
    of the frustration angle $\phi$.}
    \label{fig:ms}
\end{figure}
We can gain still more information about the evolution of the NAF state
for different couplings by calculating the sublattice magnetisation
$m_S$.  Quantum fluctuations reduce $m_S$ from its classical value $M_A
= 1/2$, and in terms of the coefficients defined above, it is given by:
\begin{equation} 
    m_S = \frac{1}{2} \sum_{\vec{q}}\left[1 
    - \frac{A_{\vec{q}}}{\sqrt{A_{\vec{q}}^2 - B_{\vec{q}}^2}}
    \right]
\end{equation}
The variation of $m_s$ as a function of $\phi$ is shown in
Figure~\ref{fig:ms}.  The sublattice magnetization of the NAF vanishes
for $\phi/\pi=0.12$, corresponding to a ratio of $J_2/J_1 = 0.38$.
This result implies that, at a semiclassical level, the NAF order
parameter is destroyed by fluctuations long before the competing CAF
becomes energetically favourable.  Historically, this was the first
signature of the existence of an intermediate, spin-gapped phase
between the NAF and CAF in the purely AF $J_1$-$J_2$ model~\cite{chandra:88}.  
We return to this point in the context of the model
with mixed FM and AF couplings below.

\subsubsection{CAF Phase}

The CAF phase can be thought of as two interpenetrating NAF lattices
of diagonal $J_2$ bonds.  At a purely classical level, these
sublattices are decoupled and can be rotated freely about one another.
However quantum fluctuations stabilise the configuration in which the
N\'eel vectors of both sublattices, and therefore the associated
spins, are collinear.  This tendency of fluctuations to favour
collinear spin configurations is well
known~\cite{shender:82,henley:89,chandra:90}, and for small $J_1/J_2$
is independant of the sign of $J_1$.  We have checked explicitly that
this order from disorder effect survives for larger, FM values of
$J_1$ by peforming spin wave calculations for more general
four--sublattice states.  Details of these will be reproduced
elsewhere.  We note that any further two--sublattice canting of the
CAF state is energeticaly unfavourable at a classical level, even for
quite large FM $J_1$.

Once the two sublattice CAF has been selected, the analysis 
of the spinwave spectrum is straightforward.
The spinwave spectrum for the CAF is once again of the form 
equation~(\ref{eqn:AB}).
For the CAF order parameter with $\vec{q}^* = (\pi,0)$ we find
\begin{eqnarray} 
    A_{\vec{q}}  &=& 2 S [2 J_2 + J_1 c_y]\\
    B_{\vec{q}}  &=& 2 S c_x [J_1 + 2 J_2 c_y]
\end{eqnarray}
in accordance with~\cite{chandra:88}
(the result for $\vec{q}^* = (0,\pi)$ can be obtained
simply by exchanging $x$ and $y$ above).
The evolution of the spin wave dispersion within the 
CAF phase is shown in Figure~\ref{fig:NAF}, plotted 
within the full BZ for the square lattice.

At the border with the NAF phase, for $\phi_c = \tan^{-1}(1/2)$
the CAF has lines of zero modes for $q_x = 0$, $q_x = \pm \pi$ and 
$q_y = \pm \pi$, but {\it not} for $q_y = 0$, and has maxima at 
$\vec{q} = (\pm \pi/2,0)$.
Within the CAF phase for AF couplings, for $\phi = \pi/4$,
the dispersion has peaked maxima 
at $\vec{q} = (\pm \pi/2,0)$,
dispersionless ridges for $q_y = \pm \pi/2$.
In all cases the dispersion had zeros 
at the wave vectors appropriate 
to the order parameter, i.e. $\vec{q} = (0,0)$ and 
$\vec{q}^* = (\pm \pi, 0)$. 
Near to these we find a linear but anisotropic 
spin wave dispersion
\begin{equation}
    \omega(\vec{q}) \sim 
    \sqrt{[v_s^x(q_x - q^*_x)]^2 + [v_s^y(q_y - q^*_y)]^2}
\end{equation}
where the spin wave velocities are given by
\begin{eqnarray}
v_S^x &=& 2J_2 + J_1 \\
v_S^y &=& \sqrt{(2J_2 + J_1)(2J_2 - J_1)} 
\end{eqnarray}
The low temperature heat capacity of the CAF is controlled by the
average spin wave velocity $\tilde{v}_S$=$(v_S^xv_S^y)^\frac{1}{2}$ and
is given by the equivalent of equation~(\ref{specNAF})
\begin{equation} 
    C_V =
    \frac{3\zeta(3)}{2\pi}\left(\frac{T}{\tilde{v}_S}\right)^2
\end{equation}
where $\zeta(3) = 1.202\ldots$.  The relevant coefficient of $T^2$, as
a function of $\phi$ is plotted in Figure~\ref{fig:cV}.

The symmetry between the x and y axes is broken by the CAF order
parameter, which is reflected in the different values of $v_x$ and
$v_y$.  However, at a semiclassical level, this symmetry breaking is
{\it not} reflected in different values the transverse susceptibility
$\chi_{\perp x}$ and $\chi_{\perp y}$ -- plotted in
Figure~\ref{fig:chi}.  This can be understood as follows -- the
transverse susceptibility associated with an AF can be expressed in
terms of the spin stiffness $\rho_s$ and the spinwave velocity $v_s$
using the hydrodynamic relation $\chi_\perp=\rho_s/v_s^2$.  However at
this level of approximation the variation of D with angle in the plane
is precisely that required to cancel the variation of $v_s$.  Once
again, the special role of the lines $J_1 = \pm 2J_2$ is evident -- at
the borders of the CAF phase the solutions for $v_y$ become imaginary,
while the transverse susceptibility diverges.

For the pure next-nearest neighbour model at $\phi = \pi/2$, the
magnetic BZ is further reduced, with additional minima at $\vec{q} =
(\pm \pi, \pm \pi)$ and $q_y = \pm \pi$.  The dispersion exhibits ridge
like maxima for $\vec{q} = (\pm \pi/2, \pm \pi/2)$.  Within the CAF
phase for partially FM coupling at $\phi = 3\pi/4$, the dispersion is
that for $\phi = \pi/4$ described above, but with $\vec{q}$ translated
by $(\pi/2, 0)$.  Finally, for $\phi_c = \pi - \tan^{-1}(1/2)$, at the
boundary with the FM state, it has maxima for $\vec{q} = (\pm \pi/2,
\pm \pi)$ and line zeros for $q_x = \pm \pi$ and $q_y = 0$, $q_y = \pm
\pi$, but {\it not} for $q_x = 0$.

While it was possible to exactly match up the spin wave dispersion
on the boundary between the NAF and FM, where the A and B
sublattices decouple, this cannot be done for the boundaries
between the CAF and NAF or CAF and FM.  The different way
in which the CAF order parameter breaks the lattice symmetry is
immediately apparent in the different number of lines of zero modes
present in the CAF for $\phi = \tan^{-1}(1/2)$ and 
$\phi = \pi - \tan^{-1}(1/2)$, as compared with the NAF or FM.

This rules out a smooth transition from one state to the
other, and at a semiclassical level, the only way in which the system
can avoid embarrassment is to dissolve the classical CAF order
parameter before the phase boundary is reached.  And indeed the
relevant sublattice magnetisation does vanish for $\phi/\pi=0.15$
($J_2/J_1 = 0.52$), as the CAF approaches the NAF, and for
$\phi/\pi=0.85$ ($J_2/J_1 = -0.51$), as the CAF approaches the FM
phase.  Consistent with this, the heat capacity (Figure~\ref{fig:cV})
shows clear anomalies as a function of $\phi$ at either end of the CAF
phase.

\begin{figure*}
    \begin{center}
	\includegraphics[width=.8\textwidth]{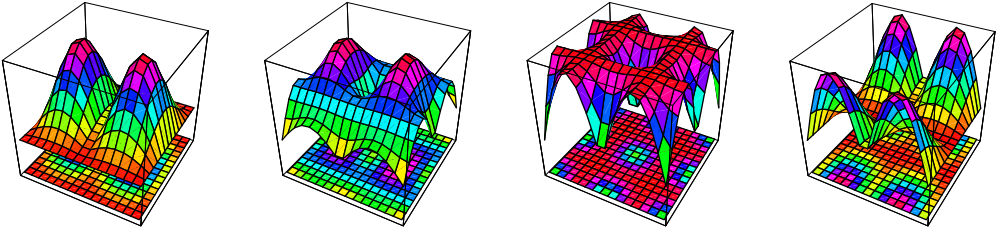}
    \end{center}
    \caption{Evolution of spinwave dispersion in CAF phase.  From left
    to right -- border with NAF, within CAF phase for AF couplings,
    pure next nearest neighbour exchange, border with FM.}
	\label{fig:CAF}
\end{figure*}

\subsection{Beyond the semiclassical picture}

The simple quantum and semiclassical arguments presented above give 
rise to an equally simple and self consistent picture of the phase 
diagram of the $J_1$-$J_2$ model as a function of $\phi$.
The model has three dominant phases, FM, NAF and CAF.  The phase
transition between the FM and NAF is straightforward. The phase
transitions between the CAF and NAF, and the CAF and FM are not,
and probably, take place through an intermediate phase.

The different estimates of the ground state energy of the system in
Figure~\ref{fig:bounds} give us further reason to believe that this
semiclassical picture is not far from the truth.  Clearly the spin wave
estimate of the ground state energy lies a long way above the lower
bound from the tetrahedral cluster.  But, as discussed in 
Appendix~\ref{low-bound}, it is possible to construct a
much better quantum bound variationally, from the numerical
diagonalization of a 9 site clusters with modified boundary
conditions.
Increasing the cluster size to
13 sites does not lead to any significant change in the ground state
energy.  Since the semiclassical estimate is within a few percent of
the quantum bound on the ground state energy for almost all values of
$\phi$, there is little room for drastic changes in the phase diagram.

But of course this is not the end of the story.  Even if our
semiclassical estimate of zero point energy is reliable, fluctuations
will strongly renormalise the semiclassical spin wave spectra and
correlation functions.  In the pure $J_1$ N\'eel AF~\cite{igarashi:92}
these corrections lead to an enhancement of the spin wave velocity
\[
    v_S\to Z_c v_S \qquad 
    Z_c=1+\frac{0.1580}{2S}+\frac{0.0216}{(2S)^2}+\ldots\approx1.1794
\]
a suppression of the perpendicular susceptibility 
%$\chi_\perp\to Z_\chi\chi_\perp$,
\[
    \chi_\perp\to
    Z_\chi\chi_\perp \qquad
    Z_\chi=1-\frac{0.551}{2S}+\frac{0.065}{(2S)^2}+\ldots\approx0.514
\]
and a slight
enhancement of the sublattice magnetisation,
\[
    m_S\to Z_mS \qquad
    Z_m=1-\frac{0.383}{2S}+\frac{0.007}{(2S)^2}+\ldots\approx0.613
\]
The values of the coefficients $Z$ as a function of $J_1$ and $J_2$ 
for $J_2 \ne 0$ remain to be calculated, and until they are known, 
quantitative comparison of the numbers found above with experiment 
must be approached with some caution.  
However experience with the pure $J_1$ model suggests that,
as long as fluctuations do not destroy magnetic order completely,
corrections are reasonably uniform and the semiclassical description of
spin correlations remains qualitatively, if not quantitatively valid.

This ``renormalised classical'' physics should be expected to break
down near the highly frustrated transitions into, and out of
the CAF phase.  The large body of existing work on the $J_1$-$J_2$ model 
with AF couplings suggests that the space between the NAF and CAF 
phases is filled by a spin-gapped `spin-liquid' phase.
Spin liquids are known to occur adjacent to a FM phase 
in the Heisenberg model on a triangular lattice with competing
FM and AF cyclic exchanges \cite{misguich:98}, and on a honey--comb lattice
with competing $J_1$, $J_2$ and $J_3$ (next--next nearest neighbour) 
interactions \cite{fouet:01}.

So what happens between the FM and CAF phases in our model ?

At least at first glance, the situation seems to be very similar.  A
high local degeneracy in the classical spectrum -- a family of
degenerate order parameters for $\phi = \pi - \tan^{-1}(1/2)$ -- leads
to line zeros in the spin wave velocity and vanishing sublattice
magnetisation exiting the CAF state towards the FM. Exactly the same
things happen at the much studied boundary with the NAF. This
similarity is by no means proof of the existence of a spin liquid state
in the $J_1$--$J_2$ model with FM $J_1$, but it is a sufficient reason
to start looking for one.

\begin{figure*}
    \begin{center}
	\includegraphics[width=.8\textwidth]{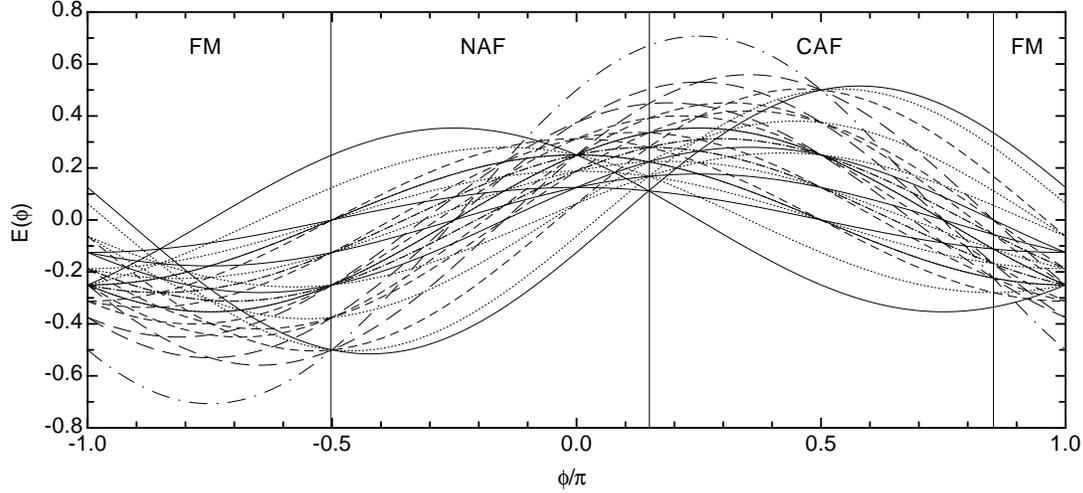}
    \end{center}
    \caption{Energy levels (per spin) of the 8 site cluster, classified
    according to total spin $\Omega$ as a function of the frustration
    angle $\phi$, in units of $J_{\rm c}$: solid lines -- $\Omega = 0$
    (singlet); dotted lines -- $\Omega = 1$ (triplet); short-dashed
    lines -- $\Omega = 2$; long-dashed lines -- $\Omega = 3$;
    dash-dotted line -- $\Omega = 4$ (maximal spin).}
    \label{fig:8site}
\end{figure*}
Further circumstantial evidence in favour of this hypothesis can be
obtained from the exact analytic diagonalization of an 8-site cluster
described in Appendix~\ref{8site}.  The resulting energy spectrum,
classified by spin, is shown in Figure~\ref{fig:8site}.  The 
straight--forward phase transition between NAF and FM states for $\phi =
-\pi/2$ shows up as a multiple crossings of ground state and excitation 
energy levels, all of which take place at the same critical
value of $\phi = -\pi/2$.

Where the singlets associated with NAF and CAF order parameters cross,
the reordering of excited states {\em does not\/} take place at a single
critical value of $\phi$, but is spread out over a finite range of
$\phi$.  Simply counting where the lowest lying triplet excitation
crosses the lowest lying singlet excitation either side of the ground
state crossing gives a remarkably good (if arbitrary) estimate of the
extent of the spin liquid region -- from $J_1/J_2 = 0.38$ to $J_1/J_2
= 0.60$, values which are comparable with those found in the existing
literature~\cite{sorella:98,melzi:01,chandra:88,dagotto:89,sushkov:01,rosner:03,misguich:03,schulz:96}.

Examining the level crossings associated with the transition from CAF
to FM we see the same extended structure.  In this case applying the
same naive criterion based on the crossing of first excitations would
predict a spin liquid region from $J_1/J_2 = -0.38$ to $J_1/J_2 =
-0.60$.  However, in order to obtain a more serious numerical estimate
of the domain of stability of the CAF and FM order parameters it would
be necessary to look at the finite size scaling of not only the first
excitation energies, but of the entire ``Anderson Tower'' of
states~\cite{anderson:52,bernu:92,azaria:93} which go to make up the
CAF order parameter in the thermodynamic limit, for a sequence of
clusters including those of relatively large sizes (e.\,g.  32 and 36
sites).  This analysis is beyond the scope of the present paper.

It also seems premature to speculate about the nature of any
possible new phase appearing between the FM and CAF. Thinking
classically, one might imagine that a spiral or canted state arises
which interpolates between the CAF and FM. Adding an AF $J_3$
interaction to the model would tend to favour such states
(cf.~\cite{chandra:90}).  However our preliminary analysis suggests
that, for the relevant range of $J_1$ and $J_2$ (with $J_3 \equiv 0$),
these states are still more unstable against fluctuations than the CAF
with which they compete.

In short, while the outcome remains uncertain, there is clearly reason 
to suspect that something interesting happens at the transition
from CAF to FM.  In marked contrast to the $J_1$--$J_2$ model
with AF interactions, the existing literature on this problem
appears to be in its infancy \cite{rastelli:86,dmitriev:97}.

\section{Finite temperature properties}
\label{finiteT}

\subsection{General considerations}

The energy and temperature scales of the $J_1$-$J_2$ model are
controlled by the single parameter $J_{\rm c}$.  In principle, this can
be determined directly from the asymptotic behaviour of the magnetic
contribution to the heat capacity at high temperatures
\cite{rosner:03,misguich:03,rosner:02}
\begin{equation}
    C_V(T\to\infty) = \frac{3}{8}\frac{J_{\rm c}^2}{T^2}+\ldots
\end{equation}
written here in units `natural' such that $k_{\rm B} = 1$.  It also
controls the deviation of the high temperature magnetic susceptibility
from a simple Curie law
\begin{equation} 
    \label{eqn:chiHT}
    \chi^{-1} (T \to \infty) = \frac{T+\Theta_{\rm CW} +
    \frac{1}{2}\frac{J_{\rm c}^2}{T} + \ldots}{C}
\end{equation}
where we again work in `natural' units such that $(g \mu_{\rm B})^2 = 1$,
\begin{equation}
C = \frac{S(S+1)}{3} = \frac{1}{4}
\end{equation}
and
\begin{equation}
    \Theta_{\rm CW} = 4(J_1 + J_2)C = J_1 + J_2 .
\end{equation}

However, while knowledge of both $J_{\rm c}$ and $\Theta_{\rm CW}$
fixes $J_1 + J_2$ and $\mid J_1 - J_2 \mid$, the sign of $J_1 - J_2$
remains undetermined since there are two possible values $\phi_{\pm}$ of the
angle $\phi$.  And because $\phi$ determines the physics of
the $J_1$-$J_2$ model, this uncertainty can lead to alternative
parameterizations of the model which lie in completely different
phases.

\begin{table*}
    \begin{center}
	\begin{tabular}{c|cc|ccccc|ccc}
	    & \multicolumn{2}{c|}{Pb$_{2}$VO(PO$_{4}$)$_{2}$}
	    & \multicolumn{5}{c|}{Li$_{2}$VOSiO$_{4}$}
	    & \multicolumn{3}{c}{Li$_{2}$VOGeO$_{4}$}\\
	    & \protect\cite{kaul:03-1} &
	    & \protect\cite{melzi:00}
	    & \protect\cite{melzi:01}
	    & \protect\cite{rosner:03}
	    & \protect\cite{misguich:03} &
	    & \protect\cite{melzi:00}
	    & \protect\cite{rosner:03}\\
	    \hline
	    $\Theta_{\rm CW} [{\rm K}]$& 4    &      &  7.4 &  8.2 & 9.65 & 7.2  &      & 5.2  & 9.8  &     \\
	    $\Theta_{\rm CW}/T_{\chi}$ & 0.49 &      & 1.39 & 1.69 &      &      &      & 1.49 \\
	    $\phi_{-}/\pi$             & 0.67 & 0.64 & 0.41 & 0.27 & 0.47 & 0.43 & 0.36 & 0.38 & 0.43 & 0.33\\
	    $\phi_{+}/\pi$             &      &-0.11 & 0.03 & 0.13 &      &      & 0.06 & 0.08 &      & 0.07\\
	    $(J_{2}/J_{1})_{-}$        &-1.64 &-2.21 &  3.5 &  1.1 & 11.7 & 4.76 & 2.13 &  2.5 & 4.76 & 1.69\\
	    $(J_{2}/J_{1})_{+}$        &      &-0.37 &  0.1 & 0.44 &      &      & 0.18 & 0.25 &      & 0.24\\
	\end{tabular}
    \end{center}
    \caption{Compilation of the experimental results and theoretical
    estimates on the Curie-Weiss temperature $\Theta_{\rm
    CW}=(J_{1}+J_{2})/k_{\rm B}$, the ratio $\Theta_{\rm CW}/T_{\chi}$
    of it to the maximum position of the uniform magnetic
    susceptibility $\chi(T)$ and the corresponding frustration
    parameters.  The data are taken
    from~\protect\cite{melzi:00,melzi:01,kaul:03-1}.  The displayed
    theoretical values for the frustration parameters obtained by fits
    to high-temperature series expansions are taken
    from~\protect\cite{rosner:03,misguich:03}.  The reference numbers
    are used to label the corresponding columns.  The unlabelled
    columns contain our own estimates derived from the dependence of
    $\Theta_{\rm CW}/T_{\chi}$ on $\phi$, see
    Figure~\protect\ref{fig:fig1}.  The $\pm$ subscripts of $\phi$ and
    $J_{2}/J_{1}$ distinguish the two different possible points in the
    $(J_{1},J_{2})$ phase diagram.}
    \label{tab:tab2}
\end{table*}
The coefficients of the high temperature series expansions of the heat
capacity and susceptibility $J_1$-$J_2$ model are known to high order
\cite{rosner:03,rosner:02}, and their reliability at low temperatures
can be greatly improved by carefully constraining the analytic
continuation of the high temperature series~\cite{roger:98,misguich:03}.  None
the less it proves very difficult to determine $J_1$ and $J_2$
unambiguously from experimental measurements of heat capacity and
magnetic susceptibility.  In the case of Li$_2$VOSiO$_4$, where both
$J_1$ and $J_2$ are believed to be antiferromagnetic, estimates of the
ratio of $J_2/J_1$ vary by more than a factor
ten~\cite{melzi:00,melzi:01,rosner:03,misguich:03,rosner:02}, see
Table~\ref{tab:tab2}.  Preliminary analysis of Pb$_2$VO(PO$_4$)$_2$,
where either $J_1$ or $J_2$ is believed to be ferromagnetic, does not
unambiguously determine which is the ferro- and which the
antiferromagnetic coupling~\cite{kaul:03-1}.

In the light of this uncertainty and controversy, we have used our
analytic solution of the 8-site cluster and an implementation of the
finite temperature Lanczos method ({\sc ftlm}) to determine the heat
capacity, uniform magnetic susceptibility and static spin structure
factor for the model on 16- and 20-site clusters with periodic
boundary conditions.  In particular, below, we present predictions for
diffuse neutron scattering cross sections which are difficult to
access by series expansion, and which can in principle remove all
ambiguity about the parameterization of the model.

Since these calculations are based on small clusters of spins, they
are of limited use in addressing questions such as the finite
temperature Ising transition observed in Monte Carlo simulations of
the classical square--lattice $J_1$--$J_2$ model~\cite{weber:03}.
However they should provide a reliable guide to physics at finite
temperatures and short length scales, particularly in the frustrated
phases of the model, where long range order is greatly suppressed.

The {\sc ftlm} is based on the Lanczos algorithm which is used to
iteratively exactly diagonalize the Hamiltonian matrix for the cluster
considered: The Lanczos algorithm starts with a randomly chosen vector
in the Hilbert space.  Successive applications of the Hamiltonian
eventually ``rotates'' the starting vector to the ground state of the
system (if the starting vector is not orthogonal to it), thereby
generating a tridiagonal matrix having eigenvalues and eigenvectors
corresponding to the lowest eigenvalues and eigenvectors of the full
Hamiltonian (this always has to be checked).  The {\sc ftlm} utilizes
the eigenvalues and eigenvectors of 1000 successive diagonalizations
with different random starting vectors to generate the partition
function and the expectation values as defined below.  For a detailed
description of it, see~\cite{jaklic:00}.

\begin{figure}
    \begin{center}
	\includegraphics[width=.8\columnwidth]{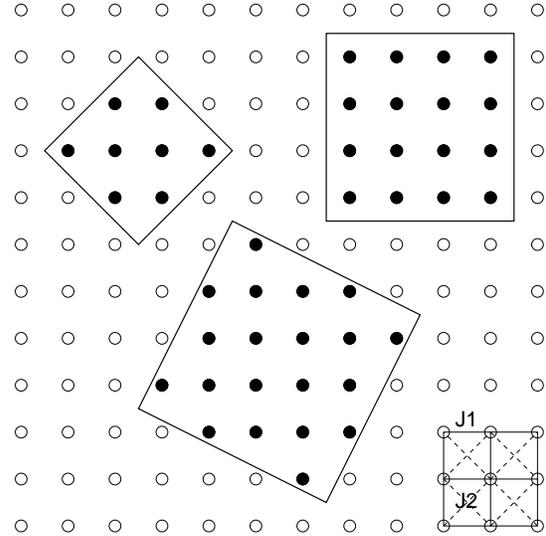}
    \end{center}
    \caption{Tiles of size eight, 16, and 20 used in the
    finite-temperature calculations.  In the lower right corner, the
    labelling of the two exchange constants is illustrated.}
    \label{fig:fig0}
\end{figure}
The three clusters, together with the realisation of the spin exchange
interactions, are shown in Figure~\ref{fig:fig0}.  Due to the symmetry
of the model, these are the only possible squares of size $N$ with
$4<N<32$ which are compatible with both collinear and N\'{e}el order.
With three cluster sizes at hand, knowing that the eight-site cluster
is almost a complete graph and the 16-site cluster a four-dimensional
hypercube, we did not attempt to perform a finite-size scaling
analysis.  For this to be meaningful, results for a system size at
least $N=32$, and as well as for different (i.\,e.~open) boundary
conditions would be needed.

\subsection{Heat capacity and magnetic susceptibility }

In units with dimensions restored, the heat capacity and the magnetic 
susceptibility are defined by
\begin{eqnarray}
    \label{eqn:chi}
    \chi(T) &=& \frac{N_{\rm A}\mu_0g^2\mu_{\rm B}^2}{Nk_{\rm B}}
    \frac{1}{T}
    \left(\left\langle\left(S_z^{\rm tot}\right)^2\right\rangle
    - \left\langle S_z^{\rm tot}\right\rangle^2\right),
    \\
    \label{eqn:cV}
    C_{V}(T) &=& \frac{N_{\rm A}}{Nk_{\rm B}}\frac{1}{T^2}
    \left(\left\langle H^2\right\rangle-\left\langle H\right\rangle^2
    \right),
\end{eqnarray}
where $\langle\dots\rangle$ denotes the thermal average, $S_{z}^{\rm
tot=\sum_{i}S_{i}^z}$ the $z$ component of the total momentum of the
system, and $N$ the number of sites of the system considered.  $N_{\rm
A}$ is the Avogadro constant, $\mu_0$ the magnetic permeability, $g$
the gyromagnetic ratio, $\mu_{\rm B}$ the Bohr magneton, and $k_{\rm
B}$ the Boltzmann constant.  For the present nonmagnetic (zero field)
case we have $\left\langle S_z^{\rm tot}\right\rangle=0$.

\begin{figure}
    \begin{center}
	\includegraphics[width=.8\columnwidth]{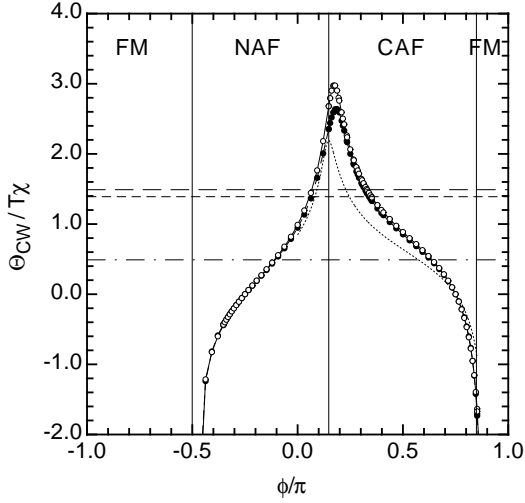}
    \end{center}
    \caption{Ratio of the Curie-Weiss temperature $\Theta_{\rm CW}$ to
    the maximum position $T_{\chi}$ of the uniform magnetic
    susceptibility $\chi(T)$ as a function of the frustration angle
    $\phi$.  The solid (dashed) line denote the results for the 20-site
    (16-site) cluster, the dotted line comprises the result for the
    eight-site cluster.  The straight horizontal lines correspond to
    the experimental values $\Theta_{\rm CW}/T_{\chi}=0.49,1.39,1.49$
    for Pb$_{2}$VO(PO$_{4}$)$_{2}$
    (dash-dotted)~\protect\cite{kaul:03-1}, Li$_{2}$VOSiO$_{4}$
    (dashed)~\protect\cite{melzi:00}, and Li$_{2}$VOGeO$_{4}$
    (long-dashed)~\protect\cite{melzi:00}, respectively.}
    \label{fig:fig1}
\end{figure}
A commonly used experimentally accessible parameter to determine the
frustration angle $\phi$ (and hence the value $J_{2}/J_{1}$) is the
ratio of the position $T_{\chi}$ of the maximum of the magnetic
susceptibility $\chi(T)$ to the Curie-Weiss temperature $\Theta_{\rm
CW}$.  To avoid singularities, we plot the inverse of this quantity,
which is shown in Figure~\ref{fig:fig1}.

Apart from the strongly frustrated spin-liquid region around
$\phi\approx\pi/6$, the differences in $\Theta_{\rm CW}/T_{\chi}$ for
the 16- and 20-site cluster are small.  This suggests that the
behaviour of $\chi(T)$, at least for $k_{\rm B}T\ge J_{\rm c}$, is
dominated by correlations which are fully taken into account already
by the small clusters, and therefore finite-size effects do not play
an important role.

A common feature of $\Theta_{\rm CW}/T_{\chi}$ for all cluster sizes is
the existence of two possible angles $\phi_{-}$ and $\phi_{+}$ for a
given value for $\Theta_{\rm CW}/T_{\chi}$, reflecting the fact that
from the knowledge of $\Theta_{\rm CW}$ and $J_{\rm c}$ alone, $\phi$
cannot be determined unambiguously.  We have indicated the experimental
values taken from~\cite{melzi:00} and~\cite{kaul:03-1} for $\Theta_{\rm
CW}/T_{\chi}$ for Pb$_{2}$VO(PO$_{4}$)$_{2}$, Li$_{2}$VOSiO$_{4}$, and
Li$_{2}$VOGeO$_{4}$ by the thin horizontal lines in
Figure~\ref{fig:fig1}.  For all three compounds, $\phi_{-}$ corresponds
to a realization of the phase with strong collinear antiferromagnetic
correlations, and $\phi_{+}$ to the phase where N\'eel-type
correlations dominate.  Interestingly, \newline some values of
$\phi_{+}$ lie very close to the spin--liquid regime \mbox{$0.115 <
\phi/\pi < 0.183$}.

Table~\ref{tab:tab2} holds a summary of the values for $\Theta_{\rm
CW}/T_{\chi}$ and
$\phi_{\pm}=\tan^{-1}\left((J_{2}/J_{1})_{\pm}\right)$ found in the
literature~\cite{melzi:00,melzi:01,kaul:03-1,rosner:03,misguich:03},
together with our findings.  $\Theta_{\rm CW}=(J_{1}+J_{2})/k_{\rm B}$
can be determined by a fit of a Curie-Weiss law to the high-temperature
tail of the susceptibility $\chi(T)$, as was done for the two Li
compounds~\cite{melzi:00,melzi:01}.  This procedure can be improved by
including higher-order terms in an expansion of $\chi(T)$, see
equation~(\ref{eqn:chiHT}).  For the Pb compound~\cite{kaul:03-1}, the
high-temperature series expansion for $\chi(T)$ found
in~\cite{rosner:03} was applied.  For the compound
Li$_{2}$VOSiO$_{4}$~\cite{melzi:00,melzi:01}, this leads to errors of
the order of 10\,\%.

In contrast to $\Theta_{\rm CW}$, the determination of the frustration
angle $\phi$ is much more involved.  Melzi et
al.~\cite{melzi:00,melzi:01} use exact diagonalization data for the
heat capacity of the eight- and 16-site
clusters~\cite{singh:90,bacci:91}, while again fits to high-temperature
series expansions are used in~\cite{kaul:03-1,rosner:03,misguich:03}.
In particular for Li$_{2}$VOSiO$_{4}$, the results for $J_{2}/J_{1}$
differ by more than an order of magnitude.  Still they indicate
qualitatively the same ordered phase in the ground state of the
compound.

However, $J_{1}$ and $J_{2}$ are not uniquely determined by this
analysis of the temperature dependence of the magnetic susceptibility
or the comparison with the behaviour of the specific heat alone.  We
will return to this issue in the next section discussing the static
spin structure factor.

\begin{figure}
    \begin{center}
	\includegraphics[width=.8\columnwidth]{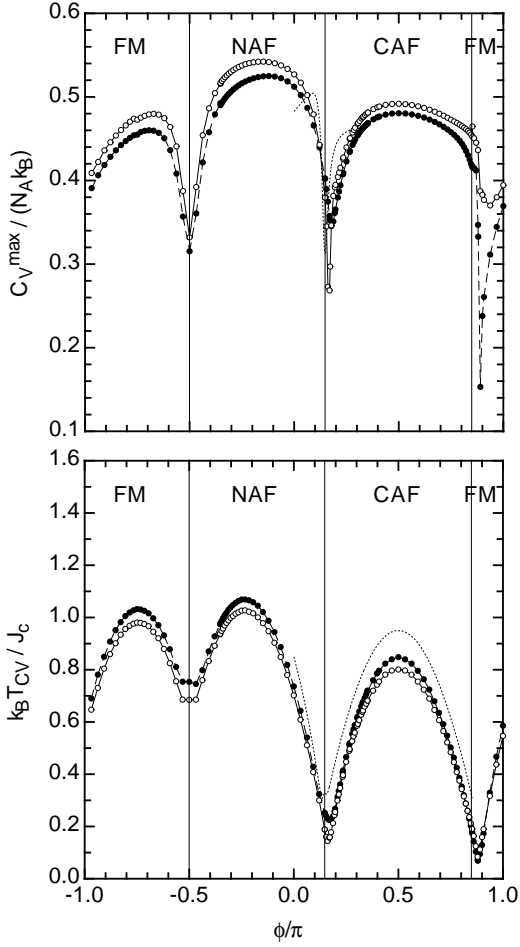}
    \end{center}
    \caption{Maximum of the heat capacity $C_{V}(T)$ and its position
    $T_{C_V}$ as functions of the frustration angle $\phi$.  The open
    (solid) circles denote the results for the 20-site (16-site)
    cluster, the dotted line denotes the eight-site cluster.}
    \label{fig:fig2}
\end{figure}
We have computed the heat capacity $C_{V}(T)$ in the full range of the
frustration angle $\phi$ for different cluster sizes.
Figure~\ref{fig:fig2} shows the maximum of the heat capacity as a
function of $\phi$.  The bottom part of the figure shows the
frustration dependence of the temperature $T_{C_V}$ at which the
maximum is reached.

Two overall effects are clearly visible: (1) Apart from the regime
with strong frustration, the maximum increases with increasing cluster
size.  (2) The maximum temperature decreases with increasing cluster
size.  Taken together, this indicates that entropy is shifted to lower
temperatures, a sign of the missing long-range correlations not
included in the partition function for the small clusters.

Our results are in qualitative agreement with those
in~\cite{melzi:01,misguich:03}.  They represent a quantitative
improvement over the estimates of~\cite{melzi:01}.  Direct comparison
with~\cite{misguich:03} is made difficult by the ambiguities associated
with analytic continuation of a series using Pad\'e approximants, and
by the fact that the limited number of cluster sizes we can use at
present do not permit a finite size scaling analysis.  In agreement
with~\cite{melzi:01}, $C_{V}^{\rm max}$ drops sharply near the
crossover between the spin liquid regime and the collinear phase around
$J_{2}/J_{1}\approx0.6$, corresponding to $\phi/\pi\approx0.17$.
Similar drops occur at the borders of the FM regime with the NAF and
CAF phases, respectively.  These drops are accompanied with a smaller
$T_{C_V}$ in order to conserve the entropy of the system.

\begin{figure}
    \begin{center}
	\includegraphics[width=.8\columnwidth]{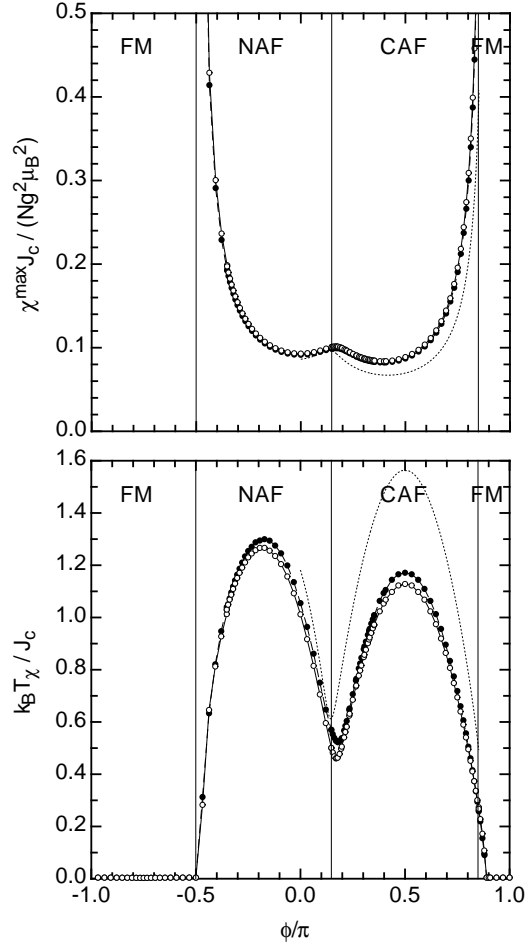}
    \end{center}
    \caption{Maximum of the uniform magnetic susceptibility $\chi(T)$
    and its position $T_{\chi}$ as functions of the frustration angle
    $\phi$.  The open (solid) circles denote the results for the
    20-site (16-site) cluster, the dotted line denotes the eight-site
    cluster.}
    \label{fig:fig3}
\end{figure}
In Figure~\ref{fig:fig3}, the behaviour of the maximum of the magnetic
susceptibility $\chi^{\rm max}$ together with the temperature at which
the maximum is reached is displayed.  In contrast to the heat capacity,
$\chi(T)$ does not display an anomaly upon crossing the spin-liquid
regime.  The maximum value diverges near the crossover to the FM
regime, while its position approaches $T=0$, which is the expected
behaviour.  Apart from that, the parameter dependence of the maximum
position $T_{\chi}$ is qualitatively the same as for $T_{C_V}$.

\begin{figure}
    \begin{center}
	\includegraphics[width=.8\columnwidth]{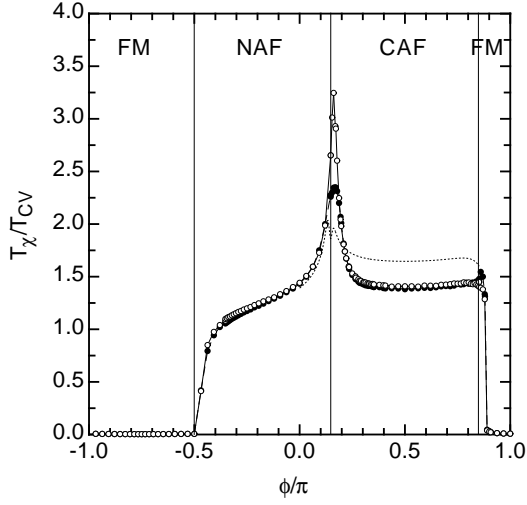}
    \end{center}
    \caption{Ratio $T_{\chi}/T_{C_V}$ of the maximum positions of the
    uniform magnetic susceptibility $\chi(T)$ and the heat capacity
    $C_{V}(T)$ as a function of the frustration angle $\phi$.  The
    solid (dashed) lines denote the results for the 20-site (16-site)
    cluster, the dotted line denotes the eight-site cluster.}
    \label{fig:fig4}
\end{figure}
Due to the sharp drop of the maximum of the heat capacity, the ratio of
the two temperatures $T_{\chi}$ and $T_{C_V}$ shows a pronounced
anomaly which has a strong dependence on cluster size in the spin
liquid regime at $\phi/\pi\approx0.17$ ($J_{2}/J_{1}\approx0.6$).  In
Figure~\ref{fig:fig4}, we have plotted $T_{\chi}/T_{C_V}$ as a function
of $\phi$.  For $0.3\le\phi/\pi\le0.85$, i.\,e., in the collinear
phase, $T_{\chi}/T_{C_V}$ does not depend on $\phi$.  For negative
values $-1/2<\phi/\pi\le-0.15$ (in the N\'eel phase),
$T_{\chi}/T_{C_V}$ depends roughly linearly on the frustration angle,
providing an additional means to determine this angle uniquely from
thermodynamic measurements alone.  (Unfortunately, none of the three
compounds discussed in this paper fall into this category.)

\subsection{Spin structure factor}

\begin{table}
    \begin{center}
	\begin{tabular}{c|cccccc}
	    {\bf q} & $(0,0)$ & $(\frac{\pi}{2},0)$ & $(\pi,0)$ &
	    $(\pi,\frac{\pi}{2})$ & $(\pi,\pi)$ &
	    $(\frac{\pi}{2},\frac{\pi}{2})$ \\
	    \hline
	    m & 1 & 4 & 2 & 4 & 1 & 4  \\
	\end{tabular} 
    \end{center}
    \caption{List of momentum vectors $\bf q$ for the 16-site square
    together with their multiplicity $m$.}
    \label{tab:kpoints}
\end{table}

The finite temperature Lanczos approach also permits the direct
evaluation of correlation functions.  We consider here the static spin 
structure factor given by
\begin{equation}
    S({\bf q},T)=\frac{1}{N}\sum_{i,j=1}^{N}e^{{\rm i}{\bf q}({\bf
    R}_{i}-{\bf R}_{j})}\left\langle{\bf S}_{i}{\bf
    S}_{j}\right\rangle.
\end{equation}
We have calculated the temperature dependence of the spin-spin
correlation functions $\left\langle{\bf S}_{i}{\bf
S}_{j}\right\rangle$ for the 16-site cluster and performed the
necessary summations to determine $S({\bf q},T)$.  The irreducible
triangle of the Brillouin zone of the 16-site cluster contains six
points which are listed together with their multiplicity in
table~\ref{tab:kpoints}.  All points lie on the edges and corners of
the triangle.

\begin{figure}
    \begin{center}
	\includegraphics[width=.8\columnwidth]{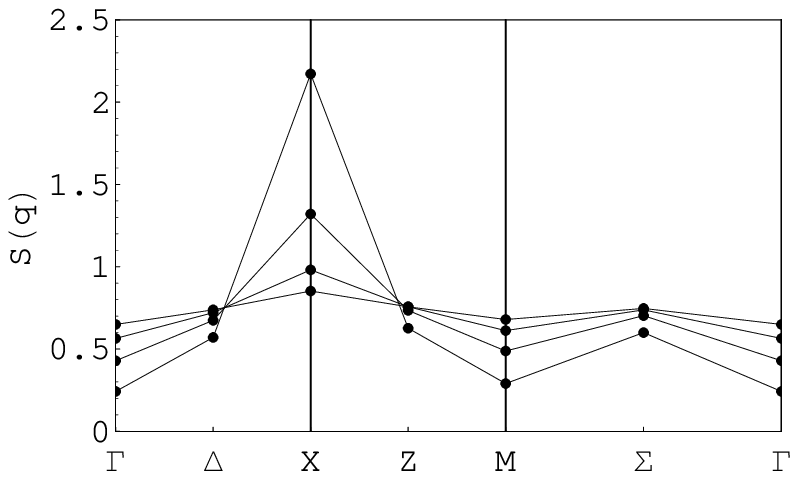}\\
	\includegraphics[width=.8\columnwidth]{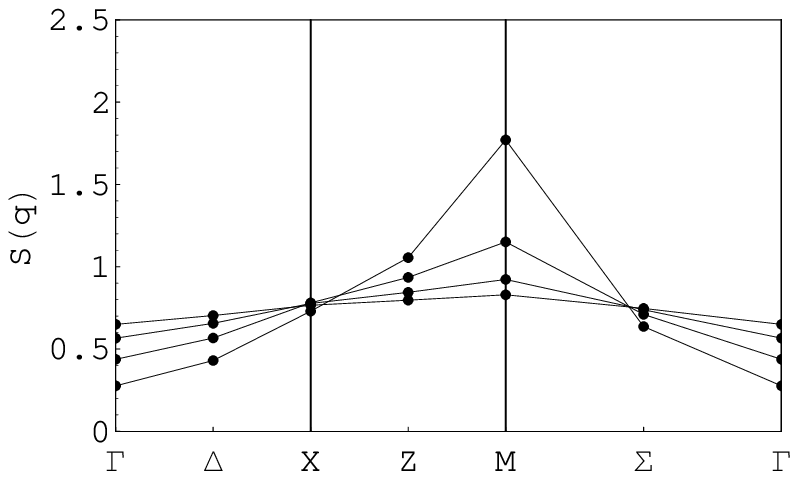}
    \end{center}
    \caption{Static spin structure factor $S({\bf q},T)$ of the 16-site
    cluster for $J_{1}/k_{\rm B}=1.25\,\rm K$, $J_{2}/k_{\rm
    B}=5.95\,\rm K$ (collinear phase, top) and $J_{1}/k_{\rm
    B}=5.95\,\rm K$, $J_{2}/k_{\rm B}=1.25\,\rm K$ (N\'eel phase,
    bottom figure).  The values chosen for $J_{1}$ and $J_{2}$
    correspond to those determined for Li$_{2}$VOSiO$_{4}$
    in~\protect\cite{misguich:03}.  The lines are guides to the eye,
    the dots denote the numerical results.  The individual curves in
    each figure, from bottom to top at ${\bf q}=\Gamma$, correspond to
    fixed temperatures $k_{\rm B}T/J_{\rm c}=1$, $2$, $4$, and $8$.}
    \label{fig:fig5}
\end{figure}
In Figure~\ref{fig:fig5}, $S({\bf q})$ is displayed as a function of
$\bf q$ lying on the edge of the irreducible Brillouin zone triangle.
We have chosen $J_{1}/k_{\rm B}=1.25\,\rm K$, $J_{2}/k_{\rm
B}=5.95\,\rm K$ (collinear phase) and $J_{1}/k_{\rm B}=5.95\,\rm K$,
$J_{2}/k_{\rm B}=1.25\,\rm K$ (N\'eel phase).  These values for $J_{1}$
and $J_{2}$ correspond to those found for Li$_{2}$VOSiO$_{4}$
in~\protect\cite{misguich:03}; the former two correspond to
$\phi_{-}/\pi=0.43$, the latter have a frustration angle of
$\phi_{+}/\pi=0.07$.  The dots represent the numerical results; the
lines connecting them are just guides to the eye.  We plot $S({\bf
q},T)$ for ten different temperatures $k_{\rm B}T/J_{\rm
c}=1,2,3\ldots10$ with an offset of one half between each two curves.

\begin{figure}
    \begin{center}
	\includegraphics[width=.8\columnwidth]{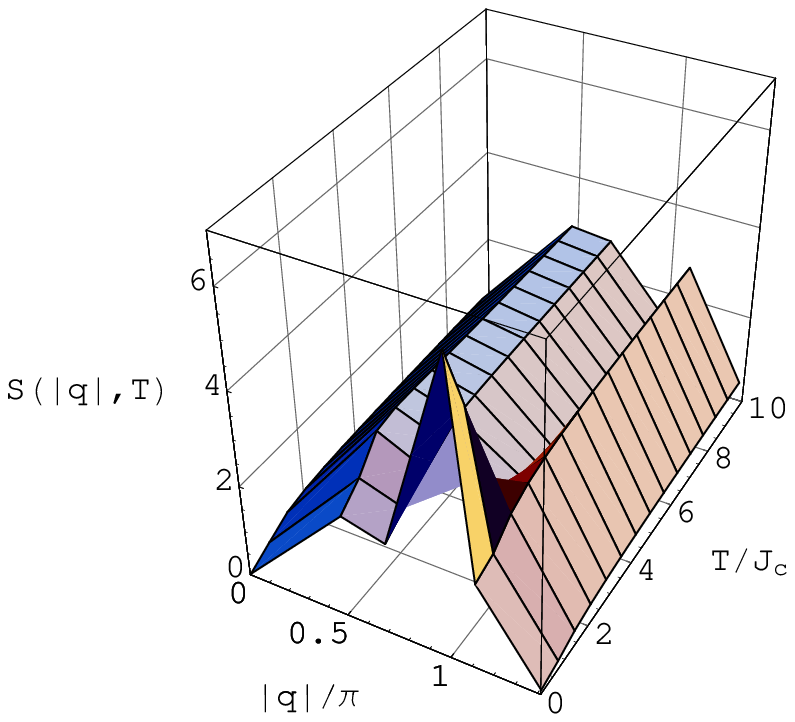}\\
	\includegraphics[width=.8\columnwidth]{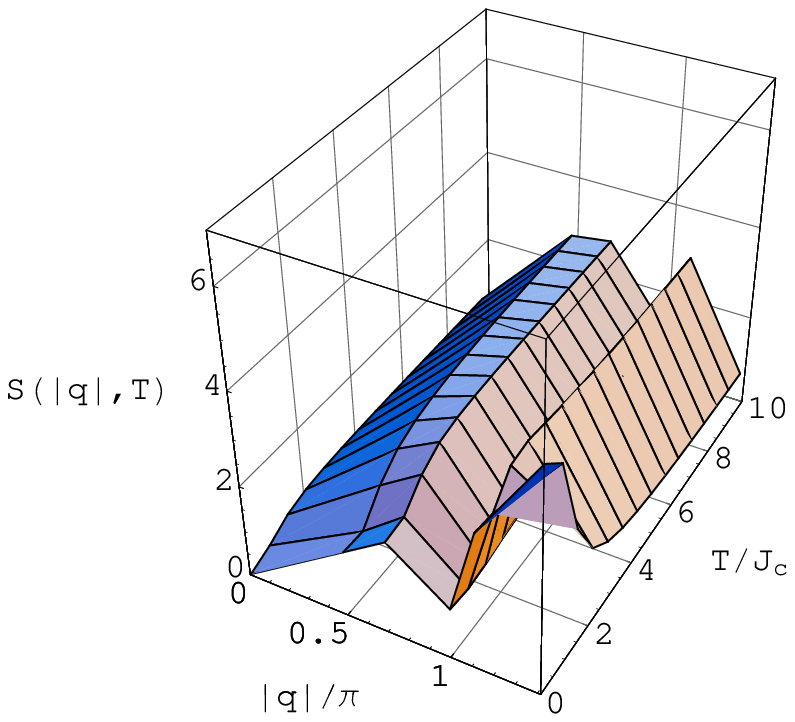}
    \end{center}
    \caption{Static spin structure factor $S(|{\bf q}|,T)$ of the
    16-site cluster for $J_{1}/k_{\rm B}=-6\,\rm K$, $J_{2}/k_{\rm
    B}=10\,\rm K$ (collinear phase, top) and $J_{1}/k_{\rm B}=10\,\rm
    K$, $J_{2}/k_{\rm B}=-6\,\rm K$ (N\'eel phase, bottom figure).  The
    values chosen for $J_{1}$ and $J_{2}$ correspond to those given for
    Pb$_{2}$VO(PO$_{4}$)$_{2}$ in~\protect\cite{kaul:03-1}.}
    \label{fig:fig6}
\end{figure}
For Pb$_{2}$VO(PO$_{4}$)$_{2}$, currently no suitable single crystals
are available to be able to measure $S({\bf q},T)$ by diffuse neutron
scattering.  Therefore, we have calculated the angular average over the
momentum transfer of $S({\bf q},T)$, which can be experimentally
determined using powder or polycrystalline material.  The results are
displayed in Figure~\ref{fig:fig6} as $S(|{\bf q}|,T)$ versus the
modulus $q=|{\bf q}|$ of the momentum transfer and the temperature $T$.
We have chosen the two frustration angles $\phi_{-}=0.64$ where the
system is in the collinear phase and $\phi_{+}=-0.11$ corresponding to
the N\'eel phase.  For the former, the maximum of $S(|{\bf q}|,T)$ is
located at $|{\bf q}|=\pi$, the latter reaches its maximum near the
zone boundary where $|{\bf q}|=\sqrt{2}\pi$.

To summarize, from Figures~\ref{fig:fig5} and~\ref{fig:fig6} we can
conclude that the structure factor provides a means to determine the
frustration ratio $\phi$ and therefore the ordering wave vector
unambiguously: For $\phi_{-}$, $S({\bf q})$ is strongly peaked at
${\bf q}=(\pi,0)$ or $(0,\pi)$, while for $\phi_{+}$ it is peaked at
${\bf q}=(\pi,\pi)$ as one would expect from the associated broken
symmetries for $\phi_\pm$ in the thermodynamic limit.

\begin{figure}
    \begin{center}
	\includegraphics[width=.8\columnwidth]{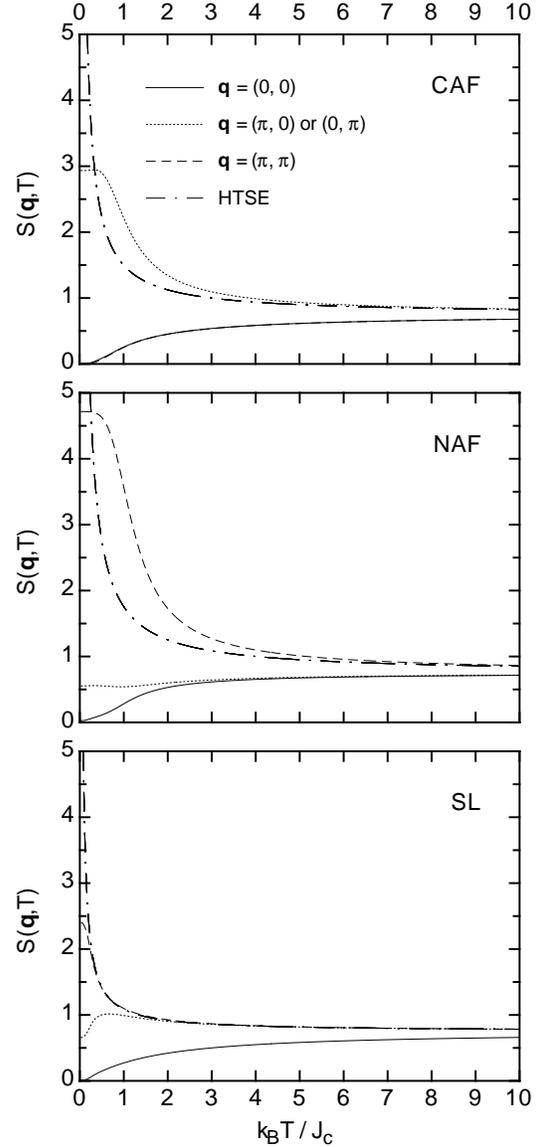}
    \end{center}
    \caption{Static spin structure factor $S({\bf q},T)$ as a function
    of temperature for ${\bf q}=(0,0)$ (solid line), ${\bf q}=(\pi,0)$
    or $(0,\pi)$ (dotted line), and ${\bf q}=(\pi,\pi)$ (dashed line)
    for three different frustration angles $\phi=\pi/2$ (collinear
    phase, top), $\phi=\tan^{-1}(-1/2)$ (N\'eel phase, middle), and
    $\phi=\tan^{-1}(1/2)$ (spin liquid, SL, phase, bottom figure).  The
    dash-dotted lines in each of the three panels denote the
    high-temperature expansions ({\sc htse}) of the structure factor at
    the respective value for $\bf q$ where $S({\bf q},T\to0)$ diverges,
    i.\,e., from top to bottom ${\bf q}=(\pi,0)$ for the collinear
    antiferromagnet, ${\bf q}=(\pi,\pi)$ for the N\'eel phase, and
    ${\bf q}=(\pi,0)\ \mbox{or}\ (\pi,\pi)$ in the spin liquid regime.}
    \label{fig:fig7}
\end{figure}
Figure~\ref{fig:fig7} displays the temperature dependence of $S({\bf
q},T)$ for different values of $\bf q$.  In the collinear and the
N\'eel phase, at low temperatures, $S({\bf q},T)$ develops a pronounced
anomaly at the ordering vector $\bf Q$ characterizing the phase.  This
anomaly is the precursor of a divergence of $S({\bf q},T\to0)$ for the
infinite system.  The asymptotic value for $k_{\rm B}T/J_{\rm
c}\to\infty$ is $S({\bf q})=S(S+1)=3/4$ in each case, as it should be.

In the collinear as well as in the N\'eel phase, the relation $S({\bf
Q})>S({\bf q})$, ${\bf q}\ne{\bf Q}$ holds for all temperatures, while
for the spin liquid regime, this is qualitatively different: Here, the
value of $S({\bf q},T)$ is approximately the same for ${\bf
q}=(\pi,\pi)$ and ${\bf q}=(\pi,0)$ or $(0,\pi)$ at temperatures
$k_{\rm B}T>J_{\rm c}$ and always larger than the value for ${\bf
q}=0$, which is an additional indicator for the strong frustration in
that phase.  The approximate equality of $S({\bf q},T)$ values for
different wave vectors also supports the picture that close to the SL
phase domains of the CAF phase may easily form inside the NAF and vice
versa as illustrated in Figure~\ref{fig:stripes}.

To illustrate our numerical findings further, we have taken the Fourier
transform of the high-temperature series expansion for the structure
factor up to first order using the result in~\cite[page
709ff]{ashcroft:88}

\begin{samepage}
    \begin{eqnarray}
	S({\bf q},T)&=&S(S+1)+\frac{S_{1}({\bf q},\phi)}{k_{\rm
	B}T/J_{\rm c}}+{\cal O}\left(\frac{1}{k_{\rm B}T/J_{\rm
	c}}\right)^{2}, \label{eqn:shtse}\\
	S_{1}({\bf
	q},\phi)&=&-\frac{zS^2(S+1)^2}{3}\left(\cos\phi\frac{1}{2}\left(\cos
	q_{x}+\cos q_{y}\right)\right.  \nonumber\\
	&\phantom{=}&\hphantom{-\frac{zS^2(S+1)^2}{3}}\left.\vphantom{\frac{1}{2}}+\sin\phi\cos
	q_{x}\cos q_{y}\right),
	\label{eqn:hats}
    \end{eqnarray}
\end{samepage}
where $z=4$ is the coordination number of the lattice.  In
Figure~\ref{fig:fig7}, the dash-dotted lines in the three panels for
the different regimes in the $(J_{1},J_{2})$ phase diagram represent
the first two terms of equation~(\ref{eqn:shtse}) for the corresponding
characteristic wave vectors ${\bf q}=(\pi,0)$ or $(0,\pi)$ in the
collinear phase, ${\bf q}=(\pi,\pi)$ in the N\'eel phase, and both of
these values in the spin liquid regime.

In the two phases where the system has a magnetically ordered ground
state, the high-temperature approximation for $S({\bf q})$ at the
respective ordering vector already underestimates the
exact-diagonalization results for temperatures $k_{\rm B}T/J_{\rm
c}\approx5$ and below.  In contrast, for the spin liquid regime,
$S({\bf q})$ is well reproduced for temperatures $k_{\rm B}T$ well
below $J_{\rm c}$, demonstrating that long-range correlation effects
are suppressed due to the presence of the strong frustration.  From
equation~(\ref{eqn:shtse}) we can also conclude that at high
temperatures $k_{\rm B}T\gg J_{\rm c}$ the relation $S({\bf
q}=(\pi,0))=S({\bf q}=(\pi,\pi))$ holds exactly for $J_{2}/J_{1}=1/2$. 
We have chosen this particular frustration ratio for the lower panel in
Figure~\ref{fig:fig7}.

\begin{figure}
    \begin{center}
	\includegraphics[width=.8\columnwidth]{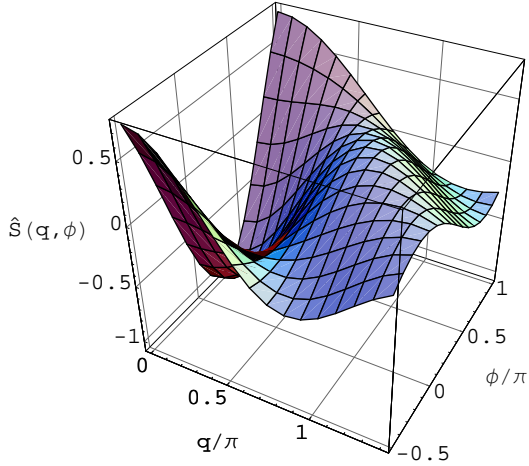}
    \end{center}
    \caption{Angular average $\hat S(q,\phi)$ of the first-order term
    of the high-temperature series expansion of the static spin
    structure factor $S({\bf q},T)$ as a function of the modulus $q$ of
    the momentum transfer and the frustration angle
    $\phi=-\pi/2\ldots\pi$.  $\phi=-\pi/2$ corresponds to $J_{1}=0$,
    $J_{2}/J_{\rm c}=-1$, which is the border between the ferromagnetic
    and the N\'eel phase in Figure~\protect\ref{fig:phasediag},
    $\phi=\pi$ corresponds to $J_{1}/J_{\rm c}=-1$ and $J_{2}=0$.
    Going from $\phi=-\pi/2$ to $\phi=\pi$, we successively scan the
    N\'eel phase, the spin liquid regime, the collinear and finally the
    ferromagnetic phase (see text).}
    \label{fig:fig8}
\end{figure}

\begin{figure}
    \begin{center}
	\includegraphics[width=.8\columnwidth]{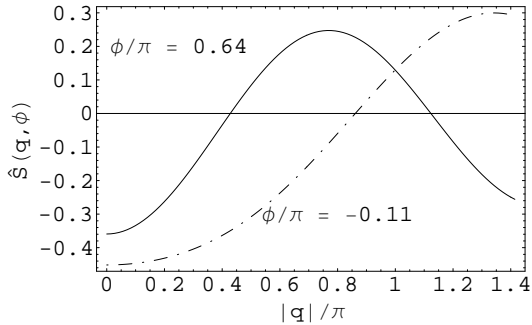}
    \end{center}
    \caption{Same as Figure~\protect\ref{fig:fig8}, here for the two
    fixed values $\phi=\phi_{\pm}$ for Pb$_{2}$VO(PO$_{4}$)$_{2}$
    (Table~\protect\ref{tab:tab2}).  Solid line:
    $\phi/\pi=\phi_{-}/\pi=0.64$ (collinear phase), dash-dotted line:
    $\phi/\pi=\phi_{+}/\pi=-0.11$ (N\'eel phase).}
    \label{fig:fig9}
\end{figure}
In Figures~\ref{fig:fig8} and~\ref{fig:fig9}, we have plotted the
angular average $\hat S(q,\phi)$ of the first-order term of the
high-temperature series expansion, equation~(\ref{eqn:hats}), given by
\begin{equation}
    \hat S(q,\phi)=\frac{1}{2\pi}\int_{0}^{2\pi}{\rm d}\alpha\,S_{1}({\bf
    q},\phi),\quad {\bf q}=q(\cos\alpha,\sin\alpha).
\end{equation}
Figure~\ref{fig:fig8} demonstrates how the maximum of this coefficient
of the structure factor evolves as a function of $\phi$.

Figure~\ref{fig:fig9} shows $\hat S(q,\phi)$ as a function of the
modulus of the momentum transfer $q=|{\bf q}|$ for the two possible
values $\phi_{\pm}$ of the frustration angle for Pb$_2$VO(PO$_4$)$_2$. 
The maximum for $\phi_{+}$ is near the zone boundary, whereas the
maximum for $\phi_{-}$ occurs at smaller values for $q$, confirming
the conclusions of our exact-diagonalization results.  Hence, for
$k_{\rm B}T\gg J_{\rm c}$, it is possible to determine experimentally
the correct value of $\phi$ from the quantity $\hat S(q,\phi)\approx
k_{\rm B}T/J_{\rm c}\left(S(q,T)-S(S+1)\right)$, and therefore
together with the temperature dependences of $\chi(T)$ and $C_{V}(T)$
the exchange parameters $J_{1}$ and $J_{2}$.

%\clearpage

\section{Summary and Conclusions}

Motivated by the discovery of Pb$_2$VO(PO$_4$)$_2$, a ``$J_1$-$J_2$
compound'' with at least one FM $J$, we have extended the
semiclassical description of the $J_1$-$J_2$ model to the case with FM
couplings.  We discussed the possible nature of the phase transitions
between the three dominant phases, FM, NAF and CAF. On the basis of
our results, the transition from CAF to FM seems to have much in
common with the transition from CAF to NAF, where an intermediate spin
liquid region is known to occur.

In addition to the phase diagram, the finite temperature properties
of the model were discussed in the light of numerical results
for 16 and 20 site clusters, using a newly implemented finite
temperature Lanczos algorithm.  In particular, we used this
to calculate spin-spin correlation functions at high temperature,
which can be compared with diffuse neutron scattering experiments.
Our numerical results should be of use in resolving the controversy
which surrounds parametrising $J_1$-$J_2$ compounds.

We also constructed a simple ``tetragonal'' mean field theory for the
$J_1$-$J_2$ model, and performed an exact {\it analytic}
diagonalization of an 8-site cluster.  While these cannot be relied
upon for quantitative comparison with experiment, they can be used to
fit susceptibility data very easily, and seem to capture the essential
physics of the model.  Both are discussed in the Appendix.

Much of the most interesting magnetic physics occurs in frustrated FM's
-- He~III on graphoil and the CMR manganites, to name but two -- and it
is our belief that the $J_1$-$J_2$ model deserves further study as
such.

\begin{acknowledgement}

It is our pleasure to acknowledge helpful conversation with Noburo
Fukushima, Christoph Geibel, George Jackeli, Enrique Kaul, R. Melzi,
Gregoire Misguich, Philippe Sindzingre and Mike Zhitomirsky.  We are
particularly grateful to Michel Roger for extensive discussions and for
providing us with the exact diagonalization code used to check the
finite temperature Lanczos results.

This work was supported by the visitors programs of the Max Planck 
Institute for the Chemical Physics of Solids and the Max Planck 
Institute for the Physics of Complex Systems, and the following 
grants: Hungarian OTKA D32689, T037451 and the EU Center of 
Excellence ICA1-CT-2000-70029.

\end{acknowledgement}

\bibliographystyle{apsrev}
\bibliography{paper}

\begin{appendix}

\section{Tetragonal Mean field theory}
\label{sec:tetrahedron}

\begin{figure}
    \begin{center}
	\includegraphics[width=\columnwidth]{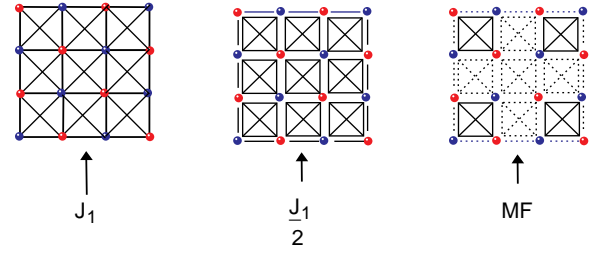}
    \end{center}
    \caption{The geometric content of the tetragonal mean-field theory
    for the $J_1$-$J_2$ model.}
    \label{fig:tetMF}
\end{figure}
\begin{table}
    \begin{center}
	\begin{tabular}{ccccc}
	    \hline
	    degeneracy & $\Omega_+$ & $\omega_A$ & $\omega_B$ & $E$ \\
	    \hline
	    1 & 2 & 1 & 1 & $J_1/2 + J_2/2$\\
	    1 & 1 & 1 & 0 & $-J_2/2$\\
	    1 & 1 & 0 & 1 & $-J_2/2$\\
	    1 & 1 & 1 & 1 & $-J_1/2 + J_2/2$\\
	    1 & 0 & 1 & 1 & $-J_1 + J_2/2$\\
	    1 & 0 & 0 & 0 & $-3J_2/2$\\
	    \hline
	\end{tabular}
    \end{center}
    \caption{The energy spectrum of a single tetrahedron.}
    \label{tab:4sitespectrum} 
\end{table}
We take as a starting point the Hamiltonian in
equation~(\ref{eqn:H4site}) for a single tetrahedron (cross-linked
square), and treat this fully connected 4-site cluster as a building
block for the lattice.  From knowledge of the spectrum and degeneracies
given in Table~\ref{tab:4sitespectrum}, we can can calculate the
partition function of a single tetrahedron exactly, and from that, its
magnetic susceptibility and heat
capacity~\cite{shannon:02}.

Starting from knowledge of the exact susceptibility of a tetrahedron
$\chi^{\rm TET}(T)$, we can construct a mean field theory for the
lattice
\begin{equation}
    \chi_{\rm MF}(T) = \frac{\chi^{\rm TET}(T)}{1 + 3(J_1 +
    J_2)\chi^{\rm TET}(T)} 
\end{equation}
The geometric content of the mean-field approximation is explained in
Figure~\ref{fig:tetMF} -- one quarter of the tetrahedra making up lattice
are treated exactly (double counting $J_1$ bonds), and the remaining
bonds are treated on a mean-field level in such a way that each spin
sees three exact and nine mean field bonds.  Because of the very high
degree of frustration involved, this approach should provide a
reasonable account of the high temperature susceptibility when the
ground state of the tetrahedron is a singlet, i.\,e.\ in the bulk NAF and
CAF phases.

At high temperatures it is possible to expand the magnetic
susceptibility found this way in $1/T$, and to make a direct comparison
with high temperature series expansions.  Only the first two terms of
equation~(\ref{eqn:chiHT}) are reproduced exactly, but the functional
form of $\chi(T)$ seems none the less to give a reasonably good account
of the experimental measured susceptibility.  Since the partition
function can be calculated for arbitrary magnetic field $h$, it is also
easy to examine non-linear effects in $\chi(T,h)$.  The 4-site cluster
is however, too small to give a convincing description of $c_V(T)$.

\section{Exact solution of the 8 site cluster}
\label{8site}

\begin{figure}
    \begin{center}
	\includegraphics[width=5truecm]{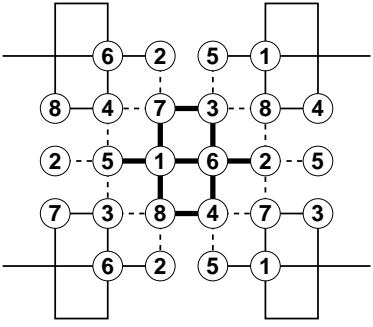}
    \end{center}
    \caption{The enumeration of sites used for the 8-site cluster.}
    \label{fig:8sitescluster}
\end{figure}
\begin{table}
    \begin{center}
	\scriptsize
	\begin{tabular}{ccccccccc}
	    \hline
	    deg.  & $S_{1\ldots8}$ & $S_{1234}$ & $S_{12}$ & $S_{34}$ &
	    $S_{5678}$ & $S_{56}$ & $S_{78}$ & $E$ \\
	    \hline
	    1 & 0 & 0 & 0 & 0 & 0 & 0 & 0 & $0$\\
	    2 & 0 & 0 & 0 & 0 & 0 & 1 & 1 & $-4J_2$\\
	    1 & 0 & 0 & 1 & 1 & 0 & 1 & 1 & $-8J_2$\\
	    4 & 0 & 1 & 0 & 1 & 1 & 0 & 1 & $-2J_1$\\
	    4 & 0 & 1 & 0 & 1 & 1 & 1 & 1 & $-2J_1-2J_2$\\
	    1 & 0 & 1 & 1 & 1 & 1 & 1 & 1 & $-2J_1-4J_2$\\
	    1 & 0 & 2 & 1 & 1 & 2 & 1 & 1 & $-6J_1+4J_2$\\
	    4 & 1 & 0 & 0 & 0 & 1 & 0 & 1 & $0$\\
	    2 & 1 & 0 & 0 & 0 & 1 & 1 & 1 & $-2J_2$\\
	    4 & 1 & 0 & 1 & 1 & 1 & 0 & 1 & $-4J_2$\\
	    2 & 1 & 0 & 1 & 1 & 1 & 1 & 1 & $-6J_2$\\
	    4 & 1 & 1 & 0 & 1 & 1 & 0 & 1 & $-J_1$\\
	    4 & 1 & 1 & 0 & 1 & 1 & 1 & 1 & $-J_1-2J_2$\\
	    4 & 1 & 1 & 0 & 1 & 2 & 1 & 1 & $-3J_1+2J_2$\\
	    1 & 1 & 1 & 1 & 1 & 1 & 1 & 1 & $-J_1-4J_2$\\
	    2 & 1 & 1 & 1 & 1 & 2 & 1 & 1 & $-3J_1$\\
	    1 & 1 & 2 & 1 & 1 & 2 & 1 & 1 & $-5J_1+4J_2$\\
	    2 & 2 & 0 & 0 & 0 & 2 & 1 & 1 & $2J_2$\\
	    2 & 2 & 0 & 1 & 1 & 2 & 1 & 1 & $-2J_2$\\
	    4 & 2 & 1 & 0 & 1 & 1 & 0 & 1 & $J_1$\\
	    4 & 2 & 1 & 0 & 1 & 1 & 1 & 1 & $J_1-2J_2$\\
	    4 & 2 & 1 & 0 & 1 & 2 & 1 & 1 & $-J_1+2J_2$\\
	    1 & 2 & 1 & 1 & 1 & 1 & 1 & 1 & $J_1-4J_2$\\
	    2 & 2 & 1 & 1 & 1 & 2 & 1 & 1 & $-J_1$\\
	    1 & 2 & 2 & 1 & 1 & 2 & 1 & 1 & $-3J_1+4J_2$\\
	    4 & 3 & 1 & 0 & 1 & 2 & 1 & 1 & $2J_1+2J_2$\\
	    2 & 3 & 1 & 1 & 1 & 2 & 1 & 1 & $2J_1$\\
	    1 & 3 & 2 & 1 & 1 & 2 & 1 & 1 & $4J_2$\\
	    1 & 4 & 2 & 1 & 1 & 2 & 1 & 1 & $4J_1+4J_2$\\
	    \hline
	\end{tabular}
    \end{center}
    \caption{\label{tab:8sitespectrum} The energy spectrum of the 8
    site cluster}
\end{table}
In Section~\ref{sec:tetrahedron}, we used the property that the energy
spectrum of a complete graph depends only on the total spin to solve a
fully connected 4-site cluster.  We can use a generalization of the
same trick~\cite{bossche:00} to solve the 8-site cluster with periodic
boundary conditions's shown in Figure~\ref{fig:8sitescluster}.  Written
in terms of complete graphs, the Hamiltonian of this cluster is:
\begin{eqnarray}
  {\cal H_{\rm 8sites}} &=& J_1 {\cal H}^{\rm CG}_{12345678}+
(2J_2-J_1) ( {\cal H}^{\rm CG}_{1234} + {\cal H}^{\rm CG}_{5678})\nonumber\\
 && -2J_2 ( {\cal H}^{\rm CG}_{12} +{\cal H}^{\rm CG}_{34} +{\cal H}^{\rm
   CG}_{56} + {\cal H}^{\rm CG}_{78})
 \label{eq:8sites}
\end{eqnarray}
where by ${\cal H}^{\rm CG}_{12345678}$ we denote the Heisenberg
Hamiltonian on a complete graph spanned by the sites 1 to 8, i.e. 
\begin{equation}
    {\cal H}^{\rm CG}_{12345678} = \sum_{i=1}^7 \sum_{j=i+1}^8 {\bf S}_i
{\bf S}_j =\frac{1}{2} \left(\sum_{i=1}^{8} {\bf S}_i \right)^2 -
\frac{1}{2}\sum_{i=1}^{8} {\bf S}_i^2 .
\end{equation}
and similarly for the others.  The nice feature of the
Hamiltonian~(\ref{eq:8sites}) is a particular hierarchy of the terms:
starting from the basic building blocks -- the states on sites
$(1,2)$, $(3,4)$, $(5,6)$, and $(7,8)$, chosen either as singlet or
triplet -- we can construct successively the definite spin states on
sites 1, 2, 3, and 4 ($S_{1234}$) and 5, 6, 7, and 8 ($S_{5678}$), and
finally on the whole cluster ($S_{12345678}$), and the energy for that
state can readily be determined.  So the problem is reduced to
bookkeeping.  The spectrum is shown in Table~\ref{tab:8sitespectrum}.

>From the knowledge of the spectrum and degeneracies, we can calculate
the exact partition function of the 8-site cluster, and from that its
heat capacity and magnetic susceptibility, just as for the 4-site
cluster.  Expanding the magnetic susceptibility obtained in this way at
high temperatures, we again find that we reproduce correctly only the
first two terms of the high temperature series expansion.  However
empirically, the magnetic susceptibility of the 8-site cluster provides
an excellent fit to experiment.  The estimate of heat capacity obtained
in this way is also more reliable than that found from the 4-site
cluster of Section~\ref{sec:tetrahedron}.

\section{Lower-bound calculation}
\label{low-bound}

\begin{figure}
  \begin{center}
      \includegraphics[width=4truecm]{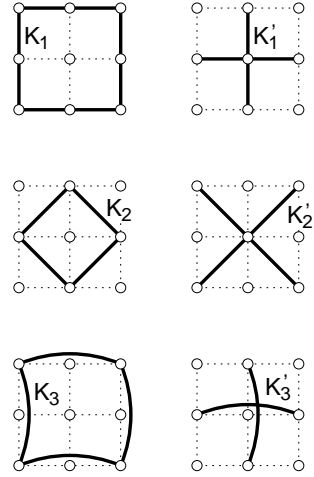}
   \end{center}
   \caption{The 9-site cluster with open boundary conditions used for
   the calculation of the lower bounds.
   \label{fig:9sites}}
\end{figure}
A lower-bound estimate of the ground-state energy for spin systems can
be obtained by exactly diagonalizing the Hamiltonian on a small cluster
with open boundary conditions.  The small cluster is chosen such that
it can be used to tile the complete lattice, in which case we can write
the Hamiltonian of the system as
\begin{equation}
    {\cal H}= \sum_i {\cal H}_i 
\end{equation}
where ${\cal H}_i$ is the Hamiltonian for an individual cluster.
The simplest such cluster for the square lattice 
with nearest neighbour bonds is the cross--linked square 
(tetrahedron), discussed above. 
The ground state wave function of the complete lattice 
\begin{equation}
    {\cal H}|\psi_0\rangle = E_0 |\psi_0\rangle
\end{equation}
can then be used as a variational wave function, providing an 
{\it upper bound} on the exact ground state energy $E_{\rm cluster}$
of a single cluster 
\begin{equation}
    \langle\psi_0 | {\cal H}_i | \psi_0 \rangle \ge E_{\rm cluster}
\end{equation}
It follows that $E_{\rm cluster}$ must in turn provide a {\it lower
bound} on the ground state energy $E_0$ per spin of the complete
lattice.  This approach was introduced by Anderson~\cite{anderson:51},
and later improved by Nishimori and Ozeki, who optimized the lower
bound by treating the exchange on the bonds within the cluster as a
variational parameter~\cite{nishimori:89}.  We have refined the method
further by allowing additional longer range bonds in the cluster which
cancel when the lattice is tiled by the sum over translated clusters.
The 9-site cluster we used is shown in Fig.~\ref{fig:9sites}, with
constraints $4K_1+2K'_1 = J_1$, $2K_2 + 2K'_2 = J_2$ and $2 K_3 + K'_3
=0$.

\end{appendix}

\end{document}